%% file: Mainsimproject.tex
\documentclass[conference]{IEEEtran}
\PassOptionsToPackage{pdfpagelabels=false}{hyperref} 

\newif\iffulledition 

\fulleditiontrue 
\usepackage[hyphens]{url}
\usepackage[colorlinks]{hyperref}
\usepackage{url}
\usepackage{fancyvrb}
\usepackage{listings}
\usepackage{amsmath}
\usepackage{booktabs}
\usepackage{epstopdf}
\usepackage{graphicx}
\usepackage{booktabs}
\usepackage{array}
\usepackage{array,multirow,graphicx}
\usepackage{pifont}
\usepackage{colortbl}
\newcommand{\cmark}{\ding{51}}
\newcommand{\xmark}{\ding{55}}
\usepackage{color}

\usepackage[font=footnotesize, caption=false]{subfig}
\usepackage{tabularx}
\newcolumntype{L}{>{\centering\arraybackslash}m{8cm}}
\usepackage{xcolor}
\definecolor{ikgreen}{RGB}{200,255,200}
\definecolor{ikred}{RGB}{255,200,200}
\definecolor{ikdarkgreen}{RGB}{100,255,100}
\definecolor{ikdarkred}{RGB}{255,100,100}
\newcolumntype{a}{>{\columncolor{ikgreen}}c}
\newcolumntype{b}{>{\columncolor{ikred}}c}

\usepackage{lipsum}

\newcolumntype{M}[1]{>{\centering\arraybackslash}m{#1}}
\newcolumntype{N}{@{}m{0pt}@{}}
\usepackage{array}
\usepackage{multirow}

\newcommand\MyBox[2]{
 \fbox{\lower0.5cm
   \vbox to 1.1cm{\vfil
     \hbox to 1.1cm{\hfil\parbox{1.1cm}{#1\\#2}\hfil}
     \vfil}%
 }%
}

\usepackage{etoolbox}
\makeatletter
\preto\lstlisting{\def\@captype{table}}
\makeatother

\makeatletter
\def\@copyrightspace{\relax}
\makeatother

\newfont{\mycrnotice}{ptmr8t at 7pt}
\newfont{\myconfname}{ptmri8t at 7pt}
\let\crnotice\mycrnotice%
\let\confname\myconfname%
\begin{document}
\lstset{language=Pascal}

\title{Towards Seamless Tracking-Free Web: Improved Detection of Trackers via One-class Learning}

\author{
\iffulledition
\IEEEauthorblockN{Muhammad Ikram$^{\dagger1}$, Hassan Jameel Asghar$^{\dagger}$,\\Mohamed Ali Kaafar$^{\dagger}$, Anirban Mahanti$^{\dagger}$}
\IEEEauthorblockA{$^{1}$ School of EET UNSW, Australia\\ $^{\dagger}$ NICTA, Australia}
\and
\IEEEauthorblockN{Balachander Krishnamurthy}
\IEEEauthorblockA{AT\&T Labs--Research, USA}
\fi
}

\IEEEoverridecommandlockouts
\makeatletter\def\@IEEEpubidpullup{9\baselineskip}\makeatother
\maketitle

\begin{abstract}

Numerous tools have been developed to aggressively block the execution of popular JavaScript programs in Web browsers. Such blocking also affects functionality of webpages and impairs user experience. As a consequence, many privacy preserving tools that have been developed to limit online tracking, often executed via JavaScript programs, may suffer from poor performance and limited uptake. A mechanism that can isolate JavaScript programs necessary for proper functioning of the website from tracking JavaScript programs would thus be useful. Through the use of a manually labelled dataset composed of 2,612 JavaScript programs, we show how current privacy preserving tools are ineffective in finding the right balance between blocking tracking JavaScript programs and allowing functional JavaScript code. To the best of our knowledge, this is the first study to assess the performance of current web privacy preserving tools.

To improve this balance, we examine the two classes of JavaScript programs and hypothesize that tracking JavaScript programs share structural similarities that can be used to differentiate them from functional {JavaScript programs}. The rationale of our approach is that web developers often ``borrow'' and customize existing pieces of code in order to embed tracking (resp. functional) JavaScript programs into their webpages. We then propose one-class machine learning classifiers using syntactic and semantic features extracted from JavaScript programs. When trained only on samples of tracking JavaScript programs, our classifiers achieve an accuracy of 99\%, where the best of the privacy preserving tools achieved an accuracy of 78\%.   
The performance of our classifiers is comparable to that of traditional two-class SVM. One-class classification, where a training set of only tracking JavaScript programs is used for learning, has the advantage that it requires fewer labelled examples that can be obtained via manual inspection of public lists of well-known trackers.

We further test our classifiers and several popular privacy preserving tools on a larger corpus of 4,084 websites with 135,656 JavaScript programs. The output of our best classifier on this data is between 20 to 64\% different from the tools under study. We manually analyse a sample of the JavaScript programs for which our classifier is in disagreement with all other privacy preserving tools, and show that our approach is not only able to enhance user web experience by correctly classifying more functional JavaScript programs, but also discovers previously unknown tracking services.

\end{abstract}

\section{Introduction}
\label{sec:introduction}
\input{introduction}

\section{Background}
\label{sec:probstatement}
\input{probstatement}

\section{Methodology}
\label{sec:methodology}
\input{methodology.tex}

\section{Effectiveness of PP-Tools}
\label{sec:epptools}
\input{epptools.tex}

\section{Classification and Validation}
\label{sec:classification}
\input{classification.tex}

\section{Evaluation in the Wild}
\label{sec:evaluation}
\input{evaluation.tex}

\section{Discussion}
\label{sec:discussion}
\input{discussion.tex}

\section{Related Work}
\label{sec:rwork}
\input{relatedwork.tex}

\section{Concluding Remarks}
\label{sec:conclusion}
\input{conclusion}

\bibliographystyle{IEEEtran}

\bibliography{simpaper.bib} 

\appendix
\section{Example of JavaScript programs Satisfying Manual Labelling Rules}
\label{appendix:label:rules}
\input{appendixa}

\end{document}

%% file: introduction.tex
Recently, several tools have been developed to aggressively block the execution of popular {JavaScript programs} in the web browser. 

JavaScript programs are frequently used to track users and tailor advertisements on websites to the browsing history and web activities
of users{~\cite{Yue:2009:, Orr:2012, Acar:2014:WNF:}}. The class of tools (including web browser plugins), developed in an attempt to preserve user privacy (e.g., NoScript~\cite{noscript}, Ghostery~\cite{ghostery}, and Adblock Plus~\cite{adblockplus}), aims to block JavaScript programs and other components of a webpage that may compromise user privacy and enable tracking. 
However, aggressive blocking can hinder proper functioning of the website and impact user's browsing experience~\cite{olejnik:hal-00747841, brokenpages} (See Appendix~\ref{sec:appendixb} for an example of how a tool blocks content necessary for proper website function.). A mechanism that can properly isolate {JavaScript programs} necessary for ``legitimate'' web functioning of the website from others that are likely to be privacy-intrusive would therefore be useful.

Web tracking in general happens through the use of numerous technologies, e.g., cookies, supercookies, Flash cookies,
ETag cookies,  HTTP referrers and {JavaScript programs}~\cite{Balaprivacyleakage:2011, roesner}.\footnote{Tracking can also be generalised to the use of
multiple user identifiers for host-tracking including IP addresses,
login IDs, web browser user-agents, etc. This is generally referred
to as stateless tracking. Interested readers may refer to
\cite{conf/ndss/YenXYYA12} for further details.} Third-party tracking, where ``unauthorised'' third parties retrieve information from the ``first party'' websites visited by users, enable a
plethora of services including analytics, advertisement and online social interactions. While third-party tracking \cite{mayer2012third} may happen through various techniques, trackers frequently use 

{JavaScript programs}{~\cite{Yue:2009:, Orr:2012, Acar:2014:WNF:,Tran:2012:TTF:}} for tracking, and as such, most tracking can be avoided by controlling the execution of  {JavaScript programs} on webpages. We refer to these {JavaScript programs} as \emph{tracking}. 
Consequently, most privacy preserving tools (PP-Tools in short) are either based on pre-defined (black)lists of URLs of third-party trackers for which the execution of JavaScript code is blocked or simply rely on blocking any third-party JavaScript program.   

However, not all {JavaScript programs} are used for tracking and many are {\it essential} for proper functioning of a website, e.g., {JavaScript programs} that enable media players to show an embedded video on a webpage. We refer to these {JavaScript programs} as \emph{functional} 
JavaScript programs. In this paper, we show that the current generation of PP-Tools are unable to achieve a balance between blocking tracking and functional {JavaScript programs}, which we believe has contributed to their poor uptake with a recent study showing about 3 to 20\% uptake among web users~\cite{pptoolusage}.

This motivates our goal to develop a machine learning classification framework that is more effective in separating tracking from functional {JavaScript programs} on a webpage. 
We impose two important design constraints. One is to avoid detection through regular expressions based on blacklists, which is used by existing PP-Tools and, as measured later in the paper, is ineffective. Another constraint is to enable classification using a small single class of {JavaScript programs} that are known to be either exclusively functional or tracking. 
Our motivation for this is that in the real-world one can expect to have knowledge of only a subset of tracking or functional {JavaScript programs}, i.e., 
we cannot expect to have an exhaustive list of either classes of programs. The key rationale of our proposed approach is that web developers often embed JavaScript code used by popular tracking libraries or re-use pieces of known JavaScript code which they customise. Likewise, code of several functional web components (including search buttons, media players embedding, shopping carts, content fetching, etc.) are generally borrowed from previously published code.    
From that perspective, our technique resembles approaches taken to detect code plagiarism and malware code signatures. 

In this paper, we develop machine learning approaches that classify functional and tracking {JavaScript programs} based on syntactic and semantic features
extracted from a number of {JavaScript programs}. We find that traditional two-class support vector machine (SVM) trained on labelled data from both functional and tracking  {JavaScript programs} can accurately
distinguish these  {JavaScript programs}. More importantly, we show that one-class machine learning classifiers, namely 
one-class SVM and positive and unlabelled (PU) learning, trained using only tracking {JavaScript programs} can achieve performance comparable to two-class SVM. 
We believe the latter approach is more practical as it requires fewer labelled samples for training which can be obtained from well-known tracking and advertising services, e.g., through blacklists of PP-Tools (although manual effort would still be required to remove wrong labels, since, as we show later, not all  {JavaScript programs} from known trackers are tracking  {JavaScript programs}). 

In summary, we make the following contributions: 

\begin{itemize}

\item We propose two machine learning approaches for automatic classification of tracking and functional  {JavaScript programs} that rely only
on partial knowledge of the former class. Instead of using static code analysis or regular expression matching on blacklists, 
we use an automated way to extract features from  {JavaScript programs} using syntactic and structural models proposed in the literature to quantitatively determine similarity between functional and tracking
 {JavaScript programs}. Our proposed approaches achieve accuracy of up to 99\%,\footnote{Accuracy is defined as the sum of true positives and negatives normalized by the population size (total number of  {JavaScript programs}). In our study, the tracking {JavaScript programs} constitute the positive class.} well above the accuracy achievable by existing
PP-Tools ($\le 78\%$) as validated through our manually labelled dataset.

\item We evaluate the effectiveness of five major PP-Tools by comparing their output against a set of 2,612 manually labelled  {JavaScript programs}
extracted from 95 different domains. Among these five, Ghostery achieves the best balance between true and false positive
rates, 0.65 to 0.08. 
Others who fare better in terms of the false positive rate ($\le 0.06$) pay the penalty with a considerably lower
true positive rate of $0.44$ or less. NoScript achieves the highest true positive rate ($0.78$), at the expense of the poorest false
positive rate ($0.21$). Our results indicate that existing PP-Tools need considerable improvement in finding an optimal balance between
true and false positives. To the best of our knowledge, this is the first study that analyses and assesses the performance of current privacy preserving tools.

\item We run our classifiers on a larger dataset of 4,084 websites, representing 135,656  {JavaScript programs} and compare their output
against the above mentioned PP-Tools to analyse their respective effectiveness in the \textit{wild}. Choosing our best classifier as a
benchmark, we observe that NoScript, Ghostery and Adblock Plus agree with the output of our classifier between 75 to 80\%, whereas
Disconnect and Privacy Badger showed an agreement of only 36\% and 43\%, respectively. Between 11\% to 14\% of the  {JavaScript programs} labelled as
functional by our classifier were contradictorily classified as tracking by the first three. The remaining two PP-Tools had this
number less than 7\%.

\item From this larger dataset we randomly sample two subsets of 100  {JavaScript programs} each, corresponding to the two sets of {JavaScript programs} labelled as functional and tracking by our classifier in disagreement with \emph{all} five PP-Tools. Through manual inspection, we found that our classifier was correct in classifying 75 out of 100  {JavaScript programs} it labelled tracking, and 81 out of 100  {JavaScript programs} it labelled functional, meaning that the PP-Tools were only correct in labelling 25 and 19  {JavaScript programs} in the two subsets, respectively. We discuss, with examples, the main reasons for misclassification by our classifier and PP-Tools. Notably, we further show how our classifier is capable of revealing previously unknown tracking services simply by relying on JavaScript code structure and semantic similarity from popular tracking services.

\end{itemize}

The rest of this paper is organised as follows. In Section \ref{sec:probstatement}, we give a background on JavaScript based web tracking and how the PP-Tools considered in our study attempt to mitigate this. We discuss our main objectives, methodology and
data collection in Section \ref{sec:methodology}. In Section~\ref{sec:epptools}, we evaluate the effectiveness of PP-Tools
using our labelled data set. Details on our choice of classifiers and their validation are described in
Section~\ref{sec:classification}. We evaluate the effectiveness of our classifiers and PP-Tools in the wild in Section~\ref{sec:evaluation}. 
In Section~\ref{sec:discussion}, we discuss possible limitations and challenges as well as avenues for improvement of our proposed scheme. We review the related work in Section~\ref{sec:rwork} and provide some concluding remarks in Section~\ref{sec:conclusion}.

%% file: probstatement.tex
\subsection{Web Tracking and JavaScript programs}
\label{subsec:webtracking}

A typical webpage consists of several web-components, e.g.,
 {JavaScript programs}, Flash-content, images, CSS, etc. When a user opens a website 
in a web browser, the fetched webpage typically generates several
other HTTP(S) connections for downloading additional components of the
webpage. These components can be downloaded from the website visited
by the user (referred to as first-party domain) or downloaded from
other third-party domains. Here, we focus on one type of web-component, namely  {JavaScript programs}, which is loaded both from first- and
third-party domains.  {JavaScript programs} are widely used by ad networks, content distribution networks (CDNs), tracking services, analytics
platforms, and online social networks (e.g., Facebook uses them to implement plugins).

\begin{figure}[!ht]
\centering
\includegraphics[width=0.45\textwidth]{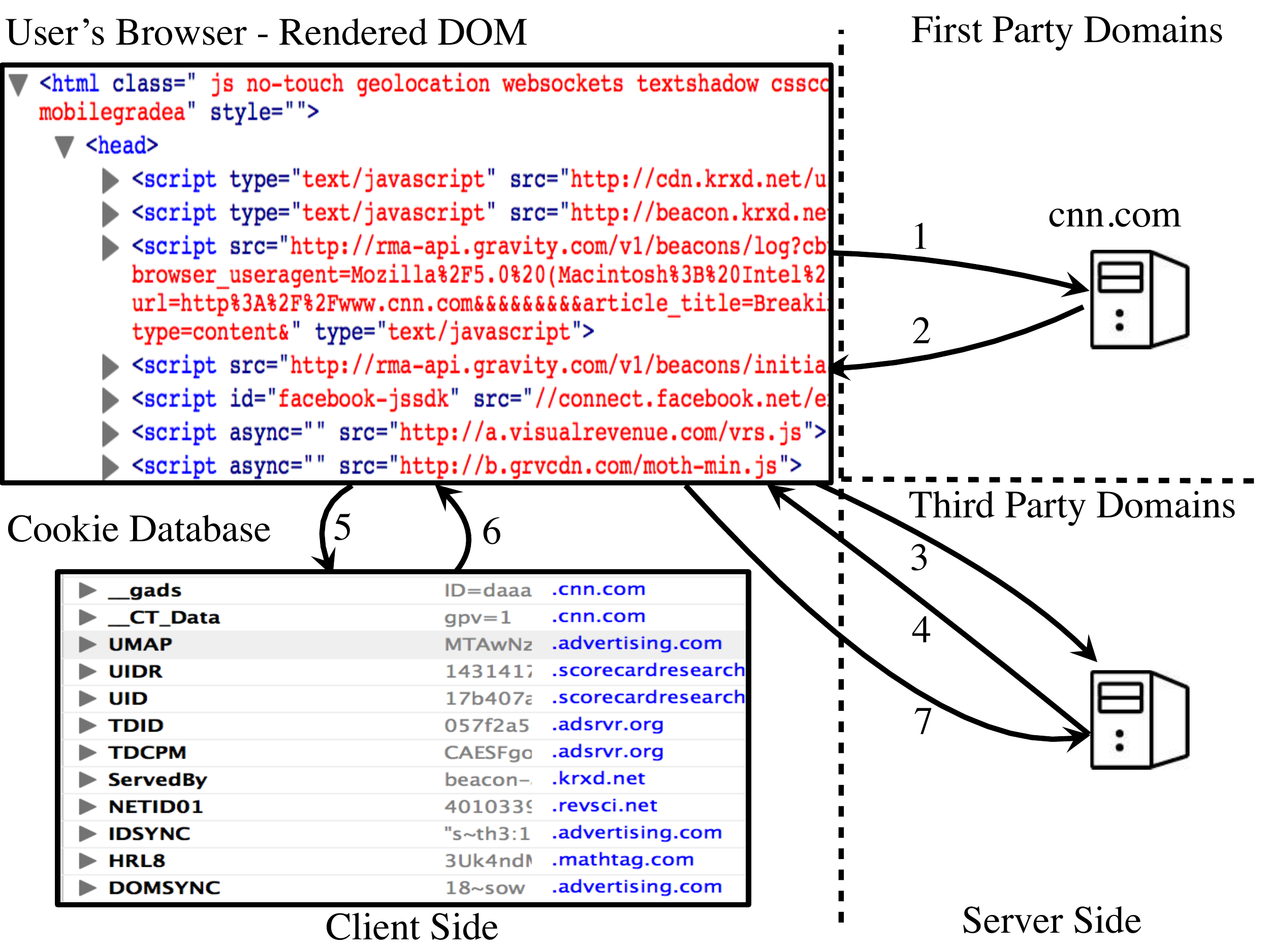}
\caption{Overview of a webpage rendering process and web tracking. Websites (in this case \texttt{cnn.com}) use third-party domains for content provisions and analytics services.} 

\label{fig:webtrackingcnn}
\end{figure}

Figure~\ref{fig:webtrackingcnn} illustrates a typical scenario of web tracking via  {JavaScript programs}. Upon fetching a webpage from first-party domains (steps 1 \& 2), the user's web browser interprets the HTML tags and executes {JavaScript programs} within the HTML \texttt{script} tags.  {JavaScript program} execution enables the web browser to send requests to retrieve additional content from third-party domains (step 3). Depending on the implemented functionalities, the  {JavaScript programs} can be considered as useful (functional), e.g., fetching content from a CDN, or as tracking. In the latter case, when the webpage is completely rendered (step 4), the {JavaScript programs} track user's activities on the webpage~\cite{TranDLJ12}, write to or read from the cookie database~\cite{olejnik:hal-00747841, Jang:2010:ESP:} (steps 5 \& 6), or reconstruct user identifiers~\cite{Acar:2014:WNF:, Nikiforakis:2013:CME}. Tracking {JavaScript programs} may also be used to transfer private and sensitive information to third-party domains (step 7)~\cite{Balaprivacyleakage:2011}.

\subsection{Privacy Preserving Tools (PP-Tools)}

\label{subsec:pptools}
In the following we briefly introduce five common PP-Tools considered in our study. 

\textbf{NoScript (NS)} blocks {JavaScript programs}, Java programs, Silverlight, Flash
and other executable content on a webpage that may undermine security including tracking \cite{noscript}. The
default behaviour is to deny, thus only allowing content that the user has explicitly permitted (whitelists). This however requires frequent user
intervention and may cause usability issues. 

\textbf{Adblock Plus (AP)} is primarily an advertisement blocking tool based on blacklists \cite{adblockplus}. It provides the option to choose from
different blacklists, e.g., \textit{EasyList}~\cite{easylist}, to
block unwanted advertisements. It searches the rendered HTML page (DOM tree) through regular expressions and blocks the
downloading of web-components such as web bugs, ads or {JavaScript programs} that belong to blacklisted tracking and advertising services.

\textbf{Disconnect (DC)} is also a blacklist based tool~\cite{disconnect}, which mainly blocks third-party tracking cookies and {JavaScript programs} from social networks such as Facebook and Twitter.

\textbf{Ghostery (GT)} finds and disables cookies and scripts that are used for tracking~\cite{ghostery}. It also searches the DOM tree through regular expressions for advertisers and trackers identified in a predefined blacklist. It provides feedback to the user to selectively unblock tracking domains. 

\textbf{Privacy Badger (PB)} uses a blacklist~\cite{pbdgerblocklist} of third-party tracking cookies that track users on multiple first-party domains. It further uses a heuristic algorithm that blocks content (JavaScript programs and ads) from third-party domains who either read high entropy cookies, or read cookies on multiple (at least 3) first-party domains, or do not comply with an acceptable ``Do Not Track" policy statement~\cite{effpolicy}.

%% file: methodology.tex
\subsection{Objectives and Overview of our Work}
\label{subsec:goalobj}

Our objectives can be summarised as follows:
\begin{enumerate}
	\item Assess the effectiveness of existing PP-Tools in terms of correctly classifying tracking and functional {JavaScript programs}.
	\item Propose an effective machine learning approach to classify tracking and functional {JavaScript programs}, trained on a subset of tracking {JavaScript programs} only. 
	\item Analyse PP-Tools and our classifier(s) in the wild and identify any trackers missed by the PP-Tools.
\end{enumerate}

For the first objective, we use a labelled dataset, composed of {JavaScript programs}, to evaluate the outcome of each of the aforementioned five PP-Tools as observed by a user navigating through a list of webpages. We used a set of 95 websites and extracted 2,612 unique {JavaScript programs} which were manually inspected and labelled as either tracking or functional according to a set of pre-identified rules (cf. Table \ref{tab:GTable}). This dataset is referred to as the \textit{labelled dataset} (cf. \S \ref{subsec:lds}). 

For the second objective, we build classifiers trained only on partial knowledge, i.e., on knowledge of labels from a subset of tracking {JavaScript programs} only. These tracking  {JavaScript programs} were extracted from the above mentioned labelled dataset. The accuracy of our classifiers is also validated using the labelled dataset, and compared against a traditional two-class classifier trained on both functional and tracking {JavaScript programs} from the labelled dataset. 

We assess the effectiveness of our classifiers and the PP-Tools \emph{in the wild}, i.e., our third objective, by further extracting 135,656  {JavaScript programs} from a set of 4,084 websites. We call this dataset, the \emph{wild dataset} (cf. \S \ref{subsec:wds}). We apply and compare results of the PP-Tools and our classifiers on this wild dataset. We also manually inspect two random samples of 100  {JavaScript programs} each on which our classifier and all the other PP-Tools disagree. As a result, we identify trackers that are missed by the PP-Tools and reveal several functional {JavaScript programs} that are being mistakenly blocked by the PP-Tools.

\subsection{Obtaining {JavaScript programs}}
\label{subsec:obtain-js}

We developed a crawler based on the Selenium webdriver~\cite{seleniumhq} for automated downloading of target webpages. The Selenium framework allows us to retrieve the whole content of a rendered webpage (DOM tree). Our crawling process is as follows: 1) use Firefox's Selenium webdriver to fetch the landing page from a given list of webpages, 2) dump the DOM tree and save it locally, 3) parse all \texttt{script} tags to find in-page {JavaScript programs}, i.e., {programs} embedded within the HTML \texttt{script} tag, and save them in files, and 4) download all external {JavaScript programs}, i.e., {programs} that are linked via external URLs. 
Due to the dynamic nature of webpages, the content of a webpage is likely to change when downloaded at different times. Moreover, webpage content may be dynamic, however our focus is JavaScript programs contained in the webpage. JavaScript programs are indeed likely to load dynamic content (ads, media content, etc.), but a JavaScript codes' nature (functional/tracking) is unlikely to change for a particular tracking domain; neither does its syntax/structure. To download the same version of a webpage for each PP-Tool, we simultaneously download webpages with PP-Tools \textit{on} and \textit{off}. 

For extracting {JavaScript programs} when the PP-Tools are turned off, referred to as PP-Tools $\textit{off}$, we obtain  {JavaScript programs} for both labelled and wild datasets as above. For the labelled dataset, these {JavaScript programs} are additionally assigned a label (tracking or functional) using Firefox and Firebug 
 as detailed in Section~\ref{subsec:lds}. Likewise, when PP-Tools are turned on (referred to as PP-Tool \emph{on}), we simultaneously obtain the set of {JavaScript programs} for each of the PP-Tools under consideration, by creating a separate Firefox profile for each tool. To unpack the compressed JavaScript codes, we use \texttt{jsbeautifier}~\cite{jsbeautifier} python library for each set of JavaScript programs when PP-Tools are \emph{on} and \emph{off}.

\subsection{The Labelled Dataset}
\label{subsec:lds}

We defined a set of 12 rules to guide our manual classification of JavaScript programs as tracking or functional. Table~\ref{tab:GTable} summarises the classification rules we used.  

\begin{table}[!th]
\begin{center}

\begin{tabularx}{\linewidth}{ c c c X  }
\toprule
Rule& JS & \# & Description\\ \midrule
R1 & \xmark & 216 & All JS that create panels and set margins for ads\\
R2 & \xmark & 115 & All JS that access and display ads \\
R3 & \xmark & 45 & All social media widgets\\

R4 & \xmark & 324 &  All in-page JS that include external JS from third-party analytics and advertisers \\

R5 & \xmark & 353 & All external JS from third-party analytics and advertisers\\
R6 & \xmark & 180 & All cookie enablers, readers or writers \\

R7 & \cmark & 542 & All external JS that provide useful functionality such as navigation menus, search and login\\
R8 & \cmark & 509 & All in-page JS that provide useful functionality \\
R9 & \cmark & 132 & All JS that fetch content from first-party content domains or third-party CDNs \\
R10 & \xmark & 103 & All JS in hidden iframe that belong to third-party analytics, advertisers and social media \\
R11 & \xmark & 40 & All JS in hidden iframe that enable, read or modify cookies \\
R12 & \cmark & 53 & All JS that track mouse or keyboard events \\
\bottomrule
\end{tabularx}
\end{center} 

 \caption{Rules for labelling  {JavaScript programs} - R stands for Rule; JS stands for JavaScript program; \# denotes the number of  {JavaScript programs} satisfying the corresponding rule in the labelled dataset; \xmark ~represents tracking  {JavaScript programs} and \cmark ~represents functional  {JavaScript programs}.}

\label{tab:GTable}
\end{table}

Rules R1 and R2 in Table~\ref{tab:GTable} label all {JavaScript programs} that create panels and set margins for ads within a webpage or the ones that fetch ads and display them, as tracking. Likewise we also label all programs that enable social media widgets, such as Facebook `Likes' and Twitter `Follow', as tracking (rule R3). We went through all the \texttt{script} tags, i.e., in-page {JavaScript programs}, and labelled them as tracking if they include an external JavaScript code (within the body) belonging to a known third-party advertiser or analytic service (R4). It follows that all \textit{external} JavaScript programs contained in a webpage that belong to a known third-party advertiser or analytics are labelled tracking (R5).

\begin {table}[!th]
\begin{center}
\tabcolsep=0.11cm
\begin{tabular}{cccccc}
\toprule
\multicolumn{6}{c}{ {JavaScript programs}} \\ 
\cline{1-6} 
 External & In-page &  Average & Total & Tracking & Functional \\
\midrule
1,353 & 1,256	& 27.5 & 2,612 & 57\% & 43\%\\
\bottomrule
\end{tabular}
\end{center}

\caption{Characteristics of  {JavaScript programs} from 95 websites in our labelled dataset.}

\label{tab:dataset:ldspptoolson1}
\end {table}

To uniquely identify users, trackers enable, read, write or update cookies on the users' machine. We used Firebug's `Cookies' panel to obtain a list of cookies. This list contains the name and ID of a cookie as well as the domain the cookie is stored for. We then searched the DOM tree by the cookie name and ID to find the JavaScript that enables, reads or modifies this cookie. Note that the JavaScript corresponding to the cookie contains its name and ID in its source code. If the domain for which the cookie is stored is a known analytic or advertiser then we mark the JavaScript as tracking; otherwise the JavaScript is marked as functional (R6). We are aware that, in principle, cookie names can be dynamically generated. However, during manual labelling we did not encounter this except for session cookies (which may not set by JavaScript programs). 

Similarly, we used Firebug's `Network' panel to identify invisible iframes that belong to third-party analytics, advertisers, or social media. We label the {JavaScript programs} belonging to these iframes as tracking {JavaScript programs} (R10). We also label the {JavaScript programs} inside iframes that enable, read, or modify cookies as tracking {JavaScript programs} (cf. rules R10 and R11). Analytics and advertisers employ hidden iframes by specifying the height and width of the iframe to zero or by positioning it so that it is out of the visible area on a webpage. A hidden iframe is positioned so that when a user interacts with a certain component of a webpage, his action and potentially the information contained in corresponding cookie(s) are redirected to the advertiser's or analytic's networks. 

All {JavaScript programs} that facilitate user interaction with the webpage's content and ensure normal functionality are labelled as functional. For instance, webpages contain  {JavaScript programs} that enable search boxes, accessibility options, authentication services, shopping carts, prompts, logins, navigation menu and breadcrumbs (rules R7 and R8). Similarly, some {JavaScript programs} are used to track mouse and keyboard events, such as right click or caps lock on or off (R1\textcolor{red}{2}) while others are used to retrieve content from either first-party content domains or third-party CDNs like Akamai (R9). We used manual list of third-party CDNs to differentiate from other content.

Note that we considered social widgets to be privacy-intrusive as they allow social networks to track users \cite{WOSN2012}; however, these could potentially be perceived as providing functional features as they allow users to interact with their social network. Notably, mouse or keyboard related JavaScript programs are only considered functional if they do not send information to third-party servers (unlike JavaScript programs that belong to e.g., `Moat'\cite{moat} that track users and send collected data to third-party servers). Likewise, JavaScript programs that tracks user's comments and send them to an external server (e.g., `Disqus'\cite{disqus}) were labelled as tracking JavaScript programs. 

We selected 95 web domains such that 50 of them were the top 50 Alexa websites, and the remaining 45 were randomly chosen from websites with Alexa rank in the range 5,000-45,000. In total, we collected 2,612  {JavaScript programs}, which consisted of 1,376 tracking and 1,236 functional  {JavaScript programs} according to our labelling. This constitutes our labelled dataset. Note that the 2,612  {JavaScript programs} obtained correspond to the PP-Tool \textit{off} setting in our terminology.  Table~\ref{tab:dataset:ldspptoolson1} summarizes the key characteristics of the {JavaScript programs} in our labelled dataset. Note that 43\% of the {JavaScript programs} are functional
indicating that the scale of the problem we are addressing is significant. Misclassifying even a small part of this can significantly impair user's web experience.

%% file: epptools.tex
We first introduce metrics used in our evaluation. 

\subsection{Aggressiveness of PP-Tools and Surrogate {JavaScript Programs}}
\label{subsec:agreesivity}

Let $H$ denote the set of DOM trees (webpages) in a given dataset (in our case, $H$ could be the labelled or the wild dataset). Let $J$ denote the set of all {JavaScript programs} from $H$. Given a DOM tree $h \in H$, $\mathsf{js}(h)$ denotes the set of {JavaScript programs} contained in $h$. This corresponds to the PP-Tools' \textit{off} setting. Let $p$ denote a given PP-Tool. Then $p(h)$ denotes the DOM tree obtained after applying $p$ to $h$, and $\mathsf{js}(p(h))$ denotes the set of {JavaScript programs} in the DOM tree $p(h)$. This corresponds to the PP-Tools' \textit{on} setting. 
We have the condition that $\mathsf{js}(p(h)) \subseteq \mathsf{js}(h)$, i.e., the set of {JavaScript programs} obtained when a PP-Tool is \textit{on} 
will always be a subset of the set of {JavaScript programs} obtained when the PP-Tool is \textit{off}. 

However, in our experiments, we found that some PP-Tools, such as Ghostery and NoScript, replace some {JavaScript programs} by new {JavaScript programs}, called \emph{surrogate} {JavaScript programs}, to ensure smooth viewing of the webpage~\cite{ssscript, ssscript1}. For instance, Ghostery replaces the JavaScript code enabling Twitter `Follow' button with a surrogate JavaScript code that displays a Ghostery button, while still blocking the functionality of the Twitter button. 
When a surrogate JavaScript codes is present in the DOM tree $p(h)$, we consider the replaced JavaScript code to be \emph{blocked} by the PP-Tool $p$, and as a result we do not include the surrogate JavaScript code in the overall count. This is consistent with our manual inspection of the surrogate lists of Ghostery and NoScript, as all the surrogates turned out to be functional 
(according to our labelling rules). 

\iffulledition
Appendix~\ref{sec:appendixb} contains more details of surrogate {JavaScript programs}. 
\fi

We define the \emph{aggressiveness} of a privacy-preserving tool $p$ as
\[
a(h) = 1 - \frac{\lvert \mathsf{js}(p(h)) \rvert}{ \lvert \mathsf{js}(h) \rvert},
\]
 which is  the fraction of {JavaScript programs} blocked by a PP-Tool $p$ from the DOM tree $h$. In settings with PP-Tools \textit{off}, aggressiveness is $0$. A PP-Tool is said to be \emph{aggressive} if it has high aggressiveness (i.e., closer to 1) for a high portion of webpages from $H$. 

\subsection{Evaluation Results}
\label{subsec:evaluation:pptoolseffectiveness}

We configured the PP-Tools so that their definitions of tracking and functional {JavaScript programs} are consistent with our labelling rules in Table~\ref{tab:GTable}. Specifically, we configured NoScript so that it does not allow iframes, consistent with rules R10 and R11 in Table~\ref{tab:GTable}. Likewise, we disabled the default option ``Allow all non-intrusive ads'' for Adblock Plus and we used The EasyList ~\cite{easylist} and Fanboy's Social Blocking List~\cite{fanboylist} as blacklists for Adblock Plus. We enabled ``Blocking all Trackers and Cookies'' option for Ghostery. We used the default settings for Privacy Badger and Disconnect since they do not provide the user with any configuration options. To account for any ``heating'' phase of the PP-Tools, we run each of the tools over another set of 100 random (non-labeled) webpages (with no overlap with the labelled dataset) prior to applying them to our labeled dataset. Summary of the set up used for PP-Tools is shown in Table \ref{tab:pptoolssettings} in Appendix~\ref{appendix:pptoolsettings}.

\begin {table}[!ht]
\begin{center}

\begin{tabular}{l*{5}{c}r}
\toprule
 & \multicolumn{5}{c}{PP-Tools On} \\ 
\cline{2-6} 
{JS}  &  NS &  GT & AP & DC & PB\\
\midrule
External   & 570	& 813	& 1,141 &	 1,206 & 1,230 \\
In-page	   & 1,118 & 	1,173 & 	1,197 & 	1,218 & 	1,208\\
Total      & 1,688 & 	1,986 & 	2,338 & 	2,424 &	 2,438 \\
Average per webpage	   & 17.6 & 20.1  &	24.4 &	25.3 &	25.4 \\
\midrule
Blocked ($\%$) & 35.38 &	24.0 &	10.5	& 7.2 & 6.7 \\
Allowed ($\%$) &  64.6 &	76.0 &	89.5 &	92.8	 & 93.3 \\				
\bottomrule
\end{tabular}

\end{center}

\caption{Characteristics of JavaScript programs (JS) from the labelled dataset marked as functional (allowed) by PP-Tools.}

\label{tab:ldspptoolson}
\end {table}

Table~\ref{tab:ldspptoolson} shows the view of {JavaScript programs} from the 95 webpages in the labelled dataset when the different PP-Tools were \textit{on}. The top-half of the table shows the number of {JavaScript programs} (in-page and external) that are allowed when a particular PP-Tool is used. The last two rows of the table show the number of {JavaScript programs} that are blocked and allowed as a percentage of the total number of {JavaScript programs} in our dataset (2,612).

\begin{figure}[!ht]

\centering
\includegraphics[width=0.40\textwidth, height=0.50\textwidth, keepaspectratio]{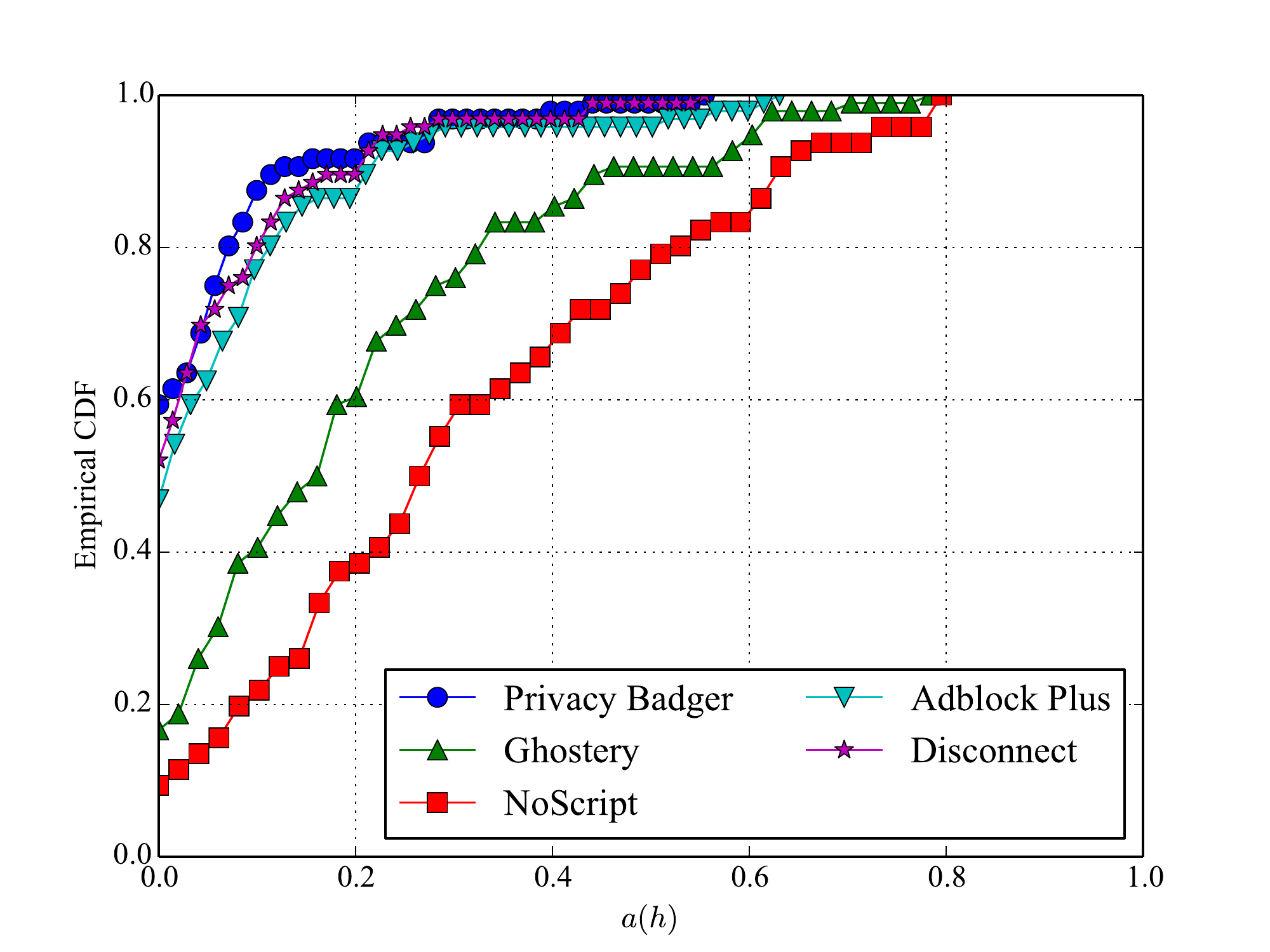}

\caption{Aggressiveness of PP-Tools.} 

\label{fig:aggressivity}
\end{figure}

We further analyse the performance of the PP-Tools by measuring their aggressiveness, i.e., $a(h)$. Figure~\ref{fig:aggressivity}
shows the cumulative distribution function of $a(h)$ for the five PP-Tools when applied on the 95 websites. We observe that NoScript's aggressiveness is more than 0.2 for about 60\% of the web domains whilst Ghostery's aggressiveness is more than 0.2 for about 40\%. In contrast, Adblock Plus, Disconnect, and Privacy Badger have an aggressiveness of more than 
0.2 for only about 10\% of domains, which indicates that they are comparatively less aggressive in blocking {JavaScript programs}. 

We are also interested to know whether there is an inverse relation between aggressiveness and effectiveness of PP-Tools, i.e., an aggressive PP-Tool breaks useful
functionality in a webpage (by incorrectly blocking functional {JavaScript programs}), and a less aggressive PP-Tool allows more tracking {JavaScript programs} go undetected. Effectiveness is defined as the balance between correctly blocking tracking {JavaScript programs} (true positives) and incorrectly blocking functional {JavaScript programs} (false positives).

\begin{table}[!thb]
\begin{center}
\resizebox{\columnwidth}{!}{  
\begin{tabular}{c|a|b|b|a}
\toprule
\multirow{2}{*}{PP-Tool} & \multicolumn{2}{c|}{Tracking} & \multicolumn{2}{c}{Functional}\\
\cline{2-5} 
& Blocked & Allowed & Blocked & Allowed \\
\hline
NoScript &  0.78	 & 0.22 &  0.21	&  0.79 \\
Ghostery & 0.65	 & 0.35 & 0.08	 & 0.92 \\ 
Adblock Plus & 0.44	 & 0.56 & 0.06	& 0.94 \\
Disconnect & 0.40	& 0.60 & 0.06	& 0.94 \\
Privacy Badger & 0.37 & 0.63 & 0.06 & 0.94\\
\bottomrule
\end{tabular}
}
\end{center}

\caption{Comparison of the output of PP-Tools against our labelled set of tracking and functional {JavaScript programs}. \textcolor{ikgreen}{\protect\rule[0pt]{2mm}{2mm}}~true positives and negatives, \textcolor{ikred}{\protect\rule[0pt]{2mm}{2mm}}~false positives and negatives.} 

\label{tab:pptoolsrecall}
\end{table}
 
For this, we measure the true positive and false positive rates of each of the PP-Tools.

Our results are shown in Table~\ref{tab:pptoolsrecall}. We find that PP-Tools' true positive rates vary from 37\% to 78\% and false positives range from 6\% to 21\%. Not surprisingly, NoScript has the highest true positive rate of 78\% at the expense of the poorest false positive rate of 21\%. Adblock Plus, Disconnect, and Privacy Badger fair better in terms of false positive rate (6\%) but pay the penalty with considerably lower true positives rates of 44\%, 40\% and 37\%, respectively. Both Ghostery and NoScript achieve the lowest average error rate (AER) of 0.215, where AER is defined as the average of false positive and negative rates. However, Ghostery is better in terms of allowing functional {JavaScript programs}, achieving a false positive rate of only 8\% with a lower true positive rate (65\%) than NoScript.

To summarise, these results suggest that current PP-Tools are ineffective in terms of striking a good balance between limiting tracking and adversely affecting useful functionalities on webpages. There are several possible reasons for this ineffectiveness. One reason could be that some of these PP-Tools collaborate with trackers and advertisers to allow some JavaScript programs that follow the PP-Tool's acceptable ads and tracking policy. For instance, on behalf of trackers and publishers, PageFair pays Adblock Plus not to block trackers and advertisements that follow their Acceptable Ads standard~\cite{Pagefair}. Another, more relevant reason for their low ineffectiveness is that current PP-Tools use rather elementary techniques such as manually maintained blacklists (whose maintenance is hard amidst the rapid growth of trackers), regular expression matching only on URLs within the \texttt{script} tag or even completely blocking the use of {JavaScript programs}. In our work, we go further by inspecting JavaScript code itself.

%% file: classification.tex
In a real setting, we cannot expect to know a priori all the possible instances of tracking and functional {JavaScript programs} due to the sheer prevalence of {JavaScript programs} on the web. Thus, realistically we may only expect to collect a small subset of {JavaScript programs} known to be functional or tracking. 
Instead of focusing on labelling both functional and tracking {JavaScript programs}, we hypothesise that it is sufficient to have partial knowledge of only the tracking JavaScript class. Our intuition is that tracking {JavaScript programs} potentially share similar characteristics and these characteristics can be leveraged in a one-class classification framework. 

In what follows, we first introduce the various models to extract JavaScript code features and then present our machine learning approaches.

\subsection{Feature Models}
\label{subsec:fselection}

\textbf{The intuition.} 
Consider the cookie setting code snippets from Google Analytics~\cite{ganalytics} and Visual Revenue~\cite{vrenue} shown as Trackers~\ref{lst:gacookiesetting} and \ref{lst:vrcookiesetting}, respectively. Notice that the two are functionally and structurally similar, with differences in variable names. More technically, the snippets result in similar canonical representations which we shall explain in Section~\ref{subsec:ngramdm}. Similar examples indicate that a similarity measure based on a feature space composed of semantic or syntactic \emph{tokens} from these {JavaScript programs} should be effective in differentiating between functional and tracking {JavaScript programs}.

\lstdefinelanguage{MyScript}{
}
\lstset{
	tabsize=2,
	basicstyle=\footnotesize\ttfamily,
	showstringspaces=false,
	showspaces=false,
	language=MyScript
}
\renewcommand{\lstlistingname}{Tracker}

\begin{lstlisting}[caption={Google Analytics Cookie Setting},frame=single, label={lst:gacookiesetting},escapeinside={@}{@}]
var _gaq = _gaq || [];
_gaq.push(['_setAccount', 'UA-1627489-1']);
_gaq.push(['_setDomainName', 'geo.tv']);
_gaq.push(['_trackPageview']);
\end{lstlisting}
\begin{lstlisting}[caption={Visual Revenue Cookie Setting},frame=single, label={lst:vrcookiesetting},escapeinside={@}{@}]
var _vrq = _vrq || [],
_vrqIsOnHP = (document.body.className || 
	'').search('pg-section') >=0 ? true : false;
_vrq.push(['id', 396]);
_vrq.push(['automate', _vrqIsOnHP]);
_vrq.push(['track', function() {}]);
\end{lstlisting}

To exemplify the existence of such similarity, we create three distinct sets of 500 JavaScript programs (tracking-only, functional-only, and tracking and functional) from our labelled dataset and calculate the \textit{term frequency with inverse document frequency} ($\mathsf{tf}\text{-}\mathsf{idf}$) based \textit{cosine} similarity values based on our feature model. Note that, being a na\"{i}ve approach, the cosine similarity metric is shown here for illustrative purposes only. We shall use our classifier later (cf.~\S\ref{subsec:classifiers}) to more efficiently classify tracking and functional JavaScript programs.

\begin{figure}[!htb]

\centering
\includegraphics[width=0.40\textwidth]{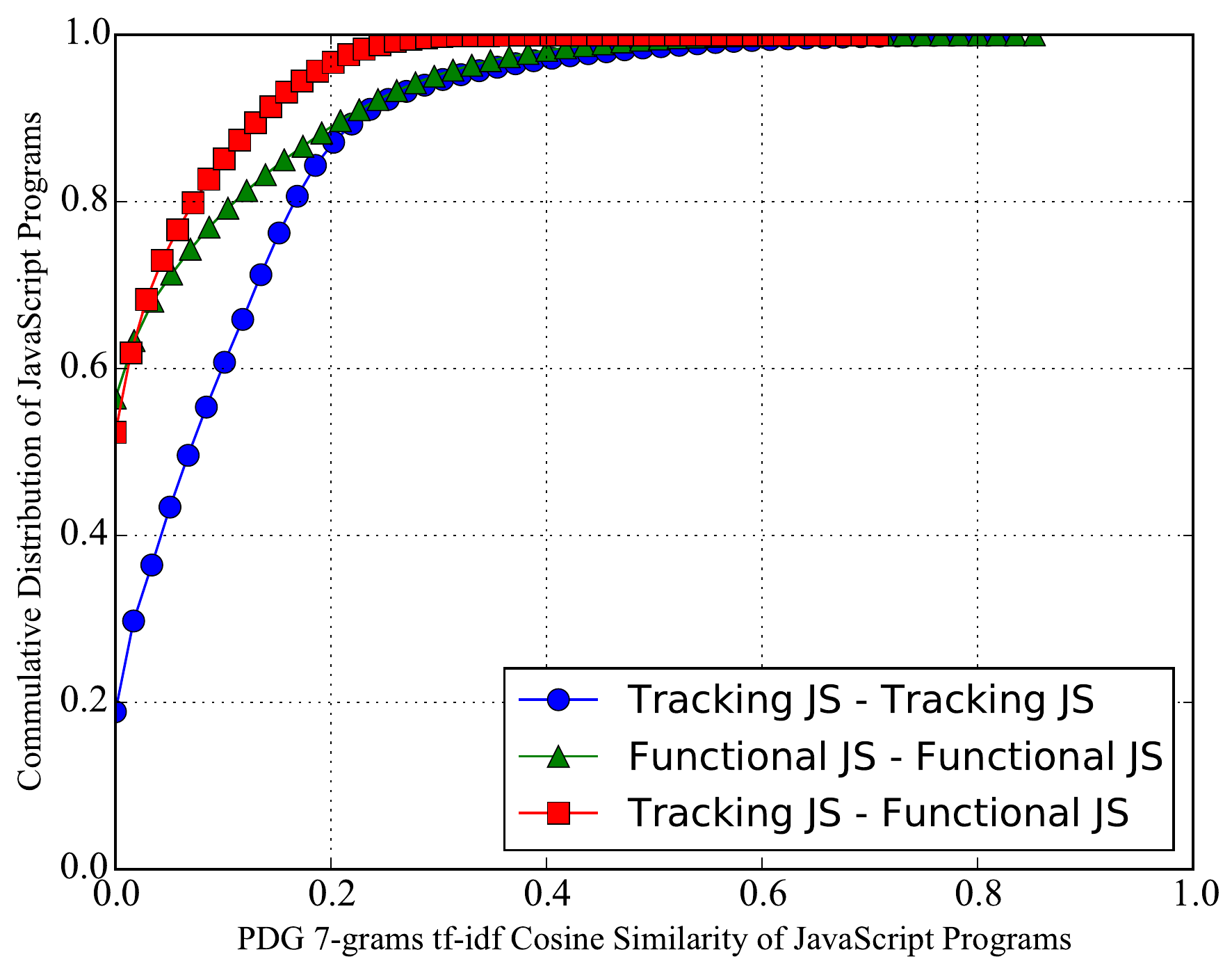}

\caption{Similarity between disjoint sets of functional and tracking JavaScript programs (JS).}

\label{fig:cscdf}
\end{figure}

Figure~\ref{fig:cscdf} plots the cumulative distribution function (CDF) of the similarity values obtained. The long tail we observe for ``tracking vs. tracking'' and ``functional vs. tracking'' as compared to ``tracking vs. functional'' suggests that tracking (resp. functional) components do indeed have higher intra-similarity. 

In Appendix~\ref{appendix:jssim}, we show further examples of tracking JavaScript programs that illustrates code similarity. 

In the rest of this section, we give details of the syntactic and semantic models in use. But first, we introduce some notation.

\noindent\textbf{Notation and Problem Formulation.} Recall that $J$ represents the set of all {JavaScript programs} in a corpus. We have a small subset $J'$ of {JavaScript programs} that are labelled. Furthermore, the labels only belong to tracking {JavaScript programs}, which we call positive labels. The totality of the negative labels, i.e., functional {JavaScript programs}, are not known beforehand. So, we have a classification problem in which we have a set $J$ whose elements can belong to two possible classes, and we only have a subset $J'$ of $J$ labelled with only one of the classes. The {JavaScript programs} from $J - J'$ can be either functional or tracking. The goal of the classifier is to obtain a correct labelling of the {JavaScript programs} in $J$. Assume that we have a feature vector $\mathbf{j}$ corresponding to some JavaScript code $j$ obtained through one of the feature extraction models, to be discussed shortly. 
Let $y$ represent the class $\mathbf{j}$ belongs to. If $\mathbf{j}$ is a tracking JavaScript, then $y = +1$; otherwise $y = -1$. Let us also introduce the label flag $l$. If $\mathbf{j}$ is labelled, i.e., is assigned a class via $y = +1$ or $-1$, then $l = 1$; otherwise $l = 0$. 

To quantitatively measure similarity between different {JavaScript programs}, we use the $\mathsf{tf}\text{-}\mathsf{idf}$ measure. Central to this measure is the term $t$. How the term $t$ is defined gives rise to the different semantic and syntactic models to be discussed shortly. For now let us assume that each JavaScript program $j \in J$ is composed of one or more terms. We then use the boolean term frequency ($\mathsf{tf}$) measure such that $\mathsf{tf}(t, j) = 1$ if $t \in j$ and $0$ otherwise. The inverse document frequency 
measure $\mathsf{idf}$ is defined as:
$\mathsf{idf}(t, j, J) = \log \frac{|J|}{\lvert \{j \in J : t \in j \} \rvert}$. 
Finally, we obtain $\mathsf{tf}$ with $\mathsf{idf}$ for each $t \in j$ over the corpus $J$, as:
$\mathsf{tf} \text{-} \mathsf{idf}(t, j, J) = \mathsf{tf}(t, j)\cdot \mathsf{idf}(t, j, J)$.
Note that, intuitively, the $\mathsf{tf}\text{-}\mathsf{idf}$ measure gives more weight to terms that are less common in $J$, since such terms are more likely to make the corresponding JavaScript program stand out. The $\mathsf{tf}\text{-}\mathsf{idf}$ thus transforms our corpus $J$ to the feature vector space. The feature vector corresponding to JavaScript program $j$ is denoted by $\mathbf{j}$. The $i$th component of $\mathbf{j}$, denoted $\mathbf{j}[i]$ corresponds to the term $t_i$ and is equal to $\mathsf{tf} \text{-} \mathsf{idf}(t_i, j, J)$. 

\subsubsection{Syntactic Model}

In the syntactic model, the term $t$ is defined as a string of characters representing \textit{one line} of JavaScript code. Note that we used \texttt{jsbeautifier} python library to unpack compressed JavaScript programs as discussed in Section~\ref{subsec:obtain-js}. Due to a large number of {JavaScript programs} contained in $J$ the dimension of the resulting feature space can become impractically large. We therefore needed to ``cap'' the feature vector space. For this we experimented with different sizes of the feature space by using the top 200, 500, 1,000, 1,500 and 2,500 terms as ranked by their $\mathsf{tf}\text{-}\mathsf{idf}$ scores, in turn. We found no considerable improvement in classification over the smallest size, i.e., 200, and since this size was also the most computationally efficient, we chose it as a cap on the feature vector space.

\subsubsection{Semantic Models}
\label{subsec:ngramdm}

In addition to the syntactic similarity between {JavaScript programs}, we are interested in a richer model that captures the structure of {JavaScript programs}. 
For this, we use the $n$-gram model introduced in \cite{Hsiao:2014:oopsala}. This approach identifies tokens in programs whose importance can then be determined through some statistical measure such as $\mathsf{tf}\text{-}\mathsf{idf}$. An example of a token is one line of JavaScript code, as used in the syntactic model. An $n$-gram, where $n \ge 1$, refers to $n$ tokens which are dependent on each other. To construct $n$-grams, the approach taken in \cite{Hsiao:2014:oopsala} converts the code into a \emph{canonical form}. 
The canonical form is an intermediate representation of a JavaScript program where variable names and loop specifics are abstracted away in an effort to apply natural language processing-like techniques on the resulting code representation. From the resultant canonical form, two different models can be derived based on how the $n$-grams are constructed. The difference lies in how different lines of the canonical code are perceived to be dependent on each other.
One model relies on sequential dependency, wherein a line of code is dependent on the execution of the previous line, and so on. 
Thus, an $n$-gram of a statement in the canonical code is a sequence of $n$ lines. We call this the \emph{sequential $n$-gram model}. 

Alternatively, since consecutive lines in source code are often unrelated to each other, instead of computing $n$-grams via this trivial ordering, program dependency graphs (PDGs) can be employed  \cite{Muchnick:1998:ACD, Hsiao:2014:oopsala}. In such a graph, a node represents a single statement from the canonical form. A node $a$ of the graph is dependent on a node $b$ if its execution depends on the result of $b$ or if it takes as input a value written by $b$. If this is true, node $a$ is connected to $b$ via a directed edge in the direction of the latter. 
An $n$-gram of a particular gram $x$ is then the subgraph of the program dependency graph consisting of all (backward) paths of length $n-1$ starting from $x$. This represents the second model and is referred to as the \emph{PDG $n$-gram model}. See Appendix~\ref{appendix:label:canonicalization} for an example of canonicalization and the resulting PDG.

Once $n$-grams have been constructed via the sequential or the PDG model, we use them in our $\mathsf{tf}\text{-}\mathsf{idf}$ measure to calculate their relative importance \cite{Hsiao:2014:oopsala}, where the term $t$ is an $n$-gram
We can have different models depending on the value of $n$ in $n$-gram. For this study, we use $4$-gram and $7$-gram variants. We also tried several values of $n$ in our $n$-gram models, but found no significant improvement beyond $n$=7.

\subsection{Machine Learning Classifiers}
\label{subsec:classifiers}

We present two machine learning approaches that apply to our problem of binary classification with only partially known instances from one of the classes used in training.

\subsubsection{One-Class SVM}

One-class SVM (OCSVM) maps the feature vectors belonging to the training data to a higher dimensional space through the use of an appropriate kernel, and then finds the hyperplane whose margin from the origin is maximized. One can view one-class SVM as a regular two-class SVM with the difference that the origin represents the only member of the second class
\cite{Scholkopf:2001:ESH}.  Given the feature vector $\mathbf{j}$ corresponding to the JavaScript program $j$, one-class SVM gives us the classifier $f(\mathbf{j}) = \langle \mathbf{w} , \mathbf{j} \rangle + b$, where the right hand side term is the equation of the hyperplane, with $\mathbf{w}$ being the vector normal to this hyperplane and $b$ being the intercept. If $f(\mathbf{j}) \ge 0$, $j$ is considered as functional; otherwise it is considered as tracking JavaScript code. 

\subsubsection{Positive and Unlabelled (PU) Learning}

The PU learning technique, translated to our problem space, constructs a probabilistic classifier that decides whether a JavaScript code is tracking
or functional from a probabilistic classifier that decides whether a JavaScript program is labelled or unlabelled. More precisely, it constructs
the classifier $f(\mathbf{j}) = \Pr [y= +1 \,|\, \mathbf{j}]$ from the classifier $g(\mathbf{j}) = \Pr [l = 1 \,|\, \mathbf{j}]$. The two
classes in $g(\mathbf{j})$ are the labelled and unlabelled {JavaScript programs}, whereas the two classes in $f(\mathbf{j})$ are positive
(tracking) and negative (functional) {JavaScript programs}.  To understand the concept behind PU learning, notice that the assumptions (a) only positive examples (tracking {JavaScript programs}) are labelled, (b) the set of labelled examples is chosen uniformly at random from all positive examples, lead to the result $\Pr[l = 1  \,|\, \mathbf{j}, y = -1] = 0$ and $\Pr[l = 1  \,|\, \mathbf{j}, y = +1] = \Pr[l = 1 \,|\, y = +1]$.
The probability $\Pr[l = 1 \,|\, y= +1]$ is the constant probability that a positive example is labelled (as it is independent of $\textbf{j}$). Now we can have $\Pr [l=1  \,|\, \textbf{j}]$ as 
\begin{align*}
\Pr [l=1  \,|\, \textbf{j}] &= \Pr [l = 1  \,|\, \textbf{j}, y = -1]\Pr[y = -1  \,|\, \textbf{j}] \\
				  &+ \Pr[l = 1  \,|\, \textbf{j}, y = +1] \Pr[y = +1  \,|\, \textbf{j}]\\
				  &= \Pr[l = 1  \,|\, y = +1] \Pr[y = +1  \,|\, \textbf{j}].
\end{align*}
As $\Pr[l = 1  \,|\, y = +1]$ is a constant, we get the classifier $f(\textbf{j})$ from $g(\textbf{j})$. We need only to estimate $\Pr[l = 1  \,|\, y = +1]$. For this, a validation set consisting of only labelled examples, say $P$, can be used. Note that, according to the assumption above, the labelled examples are all positive. Therefore, in the above equation, for $\textbf{j} \in P$, the term $\Pr[y = +1  \,|\, \textbf{j}]$ is $1$. This means that $g(\textbf{j})$ is equal to the constant $\Pr[l = 1 \,|\, y = +1]$ for the validation set $P$. Thus, we can use the \emph{trained} classifier $g(\textbf{j})$ on the validation set $P$ to estimate this constant probability by
$
\frac{1}{|P|}\sum_{\textbf{j} \in P} g(\textbf{j}).
$
In this work, we choose SVM as the trained classifier, i.e. $g(\textbf{j})$. The Reader can refer to \cite{Elkan:2008, LiuPositiveLables:2003} for a more comprehensive treatment on PU learning.

\subsection{Validation}
\label{subsec:evaluation:ecgt}

We use the traditional supervised two-class support vector machine (SSVM) \cite{cujo, Muller:2001} as a benchmark for the performance of our one-class classifiers. We run the three classifiers (one-class SVM, PU-Learning and SSVM) fed by syntactic and semantic features extracted from the {JavaScript programs} in the labelled dataset.
Syntactic features are extracted simply by treating each line of JavaScript code as an individual term. To extract semantic features, we first transform each JavaScript program into its canonical form. Then, for the sequential $n$-gram model, a term is extracted as $n$ lines preceding each line of the canonical code. For the PDG $n$-gram model, we further construct the program dependency graph from the canonical code by analysing the abstract syntax trees produced by the V8 JavaScript engine.\footnote{We used the software shared by authors of~\cite{Hsiao:2014:oopsala} to transform {JavaScript programs} into canonical forms and to construct PDGs from the canonical forms.} A term in the PDG $n$-gram model is extracted as the subgraph (consisting of paths of length $n -1$) of the PDG corresponding to each line of the canonical code. As mentioned before, feature vectors for the syntactic model were constructed by considering the top $200$ terms ranked by their $\mathsf{tf} \text{-} \mathsf{idf}$ score. On the other hand, no ``cap'' was used for the sequential $n$-gram and PDG $n$-gram models, as the feature vector size was already around 200. 

For PU-learning and one-class SVM, we use 80\% of the tracking {JavaScript programs} from the labelled dataset to constitute the training set (i.e., the training set only contains members of the positive class). We mixed the remaining 20\% of tracking {JavaScript programs} with functional {JavaScript programs} in the labelled dataset for the testing of these two classifiers. For SSVM, we use 80\% and 20\% of {JavaScript programs} (both functional and tracking) from the labelled dataset for training and testing, respectively. We empirically find the appropriate values for $\gamma$, a parameter for \textit{radial basis function} kernel \cite{Scholkopf:2001:ESH}, and $\nu$, a parameter for SVMs by performing a grid search on the ranges $2^{-15} \leq \gamma \leq 2^{0}$ and $2^{-10} \leq \nu \leq 2^{0}$ with 5-fold cross-validation on each training group. 
We use \texttt{scikit-learn}~\cite{scikit-learn}, an open source machine learning library for Python that includes a modified version of LIBSVM~\cite{Chang:2011:LibSVM}.

\subsubsection{Performance of Classifiers}

Table~\ref{tab:validationclassifiers} shows the performance of our classifiers. Note that for each feature model, we use the same training set for PU-learning and one-class SVM. We observe that, regardless of the feature model in use, PU-learning and one-class SVM exhibit similar performance. They also perform similar to supervised SVM in terms of false and true
negative rates (related to functional {JavaScript programs}). In general, except for the syntactic feature model where supervised SVM outperforms our one-class classifiers in terms of false and true positives, the three classifiers achieve very similar rates, with true positive and negative rates of up to 0.99 and false positive and negative rates of only 0.01.

\begin{table}[!th]
\begin{center}
\resizebox{\columnwidth}{!}{
\begin{tabular}{c|c|a|b|b|a}
\toprule
Feature & \multirow{2}{*}{Classifier} & \multicolumn{2}{c|}{Tracking} & \multicolumn{2}{c}{Functional}\\
\cline{3-6} 
Model& & Blocked & Allowed & Blocked & Allowed \\
\hline
Syntactic & SSVM &  0.93 & 0.07 & 0.01 & 0.99 \\
				& OCSVM & 0.88 & 0.12 & 0.02 & 0.98 \\
				& PU 		  & 0.86 & 0.14 & 0.02 & 0.98 \\
\hline 
PDG & SSVM &  0.96 & 0.04 & 0.03 & 0.97 \\
	 4-gram			& OCSVM & 0.95 & 0.05 & 0.03 & 0.97 \\
				& PU 		  & 0.93 & 0.07 & 0.04 & 0.96 \\
\hline
Sequential  & SSVM &  0.98 & 0.02 & 0.01 & 0.99 \\
	4-gram			& OCSVM & 0.98 & 0.02 & 0.02 & 0.98 \\
				& PU 		  & 0.96 & 0.04 & 0.03 & 0.97 \\
\hline
PDG & SSVM &  0.99 & 0.01 & 0.01 & 0.99 \\
	 7-gram			& OCSVM & 0.99 & 0.01 & 0.01 & 0.99 \\
				& PU 		  & 0.98 & 0.02 & 0.02 & 0.98 \\
\hline
Sequential  & SSVM &  0.99 & 0.01 & 0.01 & 0.99 \\
	7-gram			& OCSVM & 0.99 & 0.01 & 0.01 & 0.99 \\
				& PU 		  & 0.98 & 0.02 & 0.02 & 0.98 \\
\bottomrule
\end{tabular}
}
\end{center}

\caption{Performance of the classifiers against the labelled dataset of tracking and functional {JavaScript programs}. \textcolor{ikgreen}{\protect\rule[0pt]{2mm}{2mm}}~true positives and negatives, \textcolor{ikred}{\protect\rule[0pt]{2mm}{2mm}}~false positives and negatives.} 

\label{tab:validationclassifiers}
\end{table}

In comparison with the tested PP-Tools (cf.~Table~\ref{tab:pptoolsrecall}), this shows an improvement in the true positive rate by 21\% to 62\% and in the false positive rate by
5\% to 20\%.\footnote{We reiterate that our comparison is fair since we configured these PP-Tools to be consistent with our rules in Table~\ref{tab:GTable}.} Not only do our classifiers outperform the PP-Tools in effectively detecting tracking {JavaScript programs}, but they also do not suffer from high misclassification of functional {JavaScript programs} which would result in poor user web experience.

\subsubsection{Effect of Feature Models}

Table~\ref{tab:validationclassifiers} suggests that the feature models have an effect on the classification accuracy. The syntactic model has the worst performance for all three classifiers. The PDG 7-gram and sequential 7-gram models in contrast show the best results for all the classifiers. We improve the false negative rate by 11-12\% in the case
of the one-class classifiers by using the 7-gram models. Interestingly, the performance of the classifiers for classifying functional {JavaScript programs} is similar across all
feature models, which suggests that perhaps functional {JavaScript programs} have more inter-similarity than the inter-similarity between tracking {JavaScript programs}.

{Our results show that one-class classifiers, as non-expensive learning techniques, perform similarly}. Now we aim to apply our one-class SVM and PU-learning classifiers in the wild and compare their output to PP-Tools. For this purpose, we choose the best and the worst performers of the lot: sequential 7-gram and the syntactic model, respectively. Note that although, performance wise, sequential $7$-gram and PDG $7$-gram are similar, we chose the former as it requires less pre-processing (no construction of PDGs).

%% file: evaluation.tex
We evaluate and compare the output of our classifiers and the PP-Tools on a set of {JavaScript programs} collected from a large number of websites, called the \textit{wild dataset}.  

\subsection{The Wild Dataset}
\label{subsec:wds}

The wild dataset consists of {JavaScript programs} extracted from the landing page of 4,084 websites. These websites were selected such that 
3,500 of them were top Alexa websites ranked between 51 to 3,550 inclusive (since the top 50 websites were already used in the labelled dataset), with the remaining 584 having a rank in excess of 5,000 
(excluding the 45 websites with rank in this range used in the labelled dataset (cf. \S\ref{subsec:lds})). A total of 135,656 {JavaScript programs} were present in these websites. 
The composition of {JavaScript programs} into external and in-page {JavaScript programs} is shown in Table~\ref{tab:wdspptoolson} 
(under the column labelled PP-Tools \emph{off}). The table also shows the number of in-page and external {JavaScript programs}
allowed by each PP-Tool considered in our work. Again, we observe that NoScript is the most aggressive tool, blocking on average 37.7\% of the {JavaScript programs} per website, closely followed by Ghostery (34.8\%) and AdBlock Plus (26\%). In comparison, Privacy Badger and Disconnect block 11.5\% of {JavaScript programs} per website.  
  
\begin{table}[!thb]
\begin{center}
\resizebox{\columnwidth}{!}{
\begin{tabular}{ l c c c c c c}
\hline
  &  PP-Tools & \multicolumn{5}{c}{PP-Tools \emph{on}} \\
\cline{3-7}
JS & \emph{off} & NS &  GT & AP & DC & PB\\
\hline
External  &     71,582  & 29,345 &      38,492 &        48,191 &        59,488 &        60,817 \\
In-page & 64,074        & 54,972        & 49,952 &      51,389  & 60,546        & 59,250\\

Total &         135,656 & 84,428        & 88,444        & 99,580        & 120,034        & 120,067 \\
Average per webpage & 33.2   & 20.7         & 21.7  & 24.4  & 29.4  & 29.4 \\
\hline
Blocked (\%) & -   & 37.7   & 34.8  & 26.6  & 11.5  & 11.5 \\
Allowed (\%) & -   & 63.3 & 65.2  & 73.4  & 88.5  & 88.5 \\

\hline
\end{tabular}
}
\end{center}

\caption{Characteristics of {JavaScript programs} (JS) collected from 4,084 websites with PP-Tools \textit{on} and \textit{off} as viewed from a Firefox Selenium-controlled browser.}

\label{tab:wdspptoolson}
\end{table}

\subsection{Comparing PP-Tools and our Classifiers}
\label{subsec:evaluation:wds}

We trained our PU-Learning and one-class SVM classifiers using only the tracking {JavaScript programs} from the labelled dataset. 
The trained classifiers were then run on the wild dataset.
Next, we assess the extent to which our two classifiers agree or disagree with each PP-Tool.

\subsubsection{Agreement Ratio}

Denote by $T_c$ and $F_c$ the set of tracking and functional {JavaScript programs}, respectively,
in the wild dataset as classified by a classifier $c$ (i.e., one-class SVM or PU learning). Likewise, $T_p$, resp. $F_p$, denotes tracking, resp. functional, JavaScript programs as marked by the PP-Tools $p$.

Then the ratio of agreement between $c$ and $p$ on tracking {JavaScript programs} is ${| T_c \cap T_p |}/{|J|}$, where $J$ is the set of {JavaScript programs} in the wild dataset.
Similarly, the ratio of agreement on the functional {JavaScript programs} is ${| F_c \cap F_p |}/{|J|}$. 
The ratio of agreement between a classifier $c$ and a PP-Tool $p$ is simply ${(| T_c \cap T_p | + | F_c \cap F_p |)}/{|J|}$.
To avoid excessive notation, we shall denote the ratios simply by their constituent sets. For instance, the agreement ratio for tracking {JavaScript programs} will simply be written as $T_c \cap T_p$. Table~\ref{tab:wds:agreement} lists the values of agreement and the corresponding disagreement ratios (defined as 1 minus the agreement ratio). Figure~\ref{fig:ocsvm:agreement} in Appendix~\ref{subsection:ex_analyzing_disagreement} illustrates these ratios graphically.

\begin{table*}[!ht]
\begin{center}
\begin{tabular}{c|c|c|a|b|b|a|c|c}
\hline
Feature Model & Classifier & PP-Tool & $T_c \cap T_p$ & $T_c \cap F_p$ & $F_c \cap T_p$ & $F_c \cap F_p$ & Agreement & Disagreement\\
\hline
Syntactic & OCSVM & NoScript & 0.56 & 0.10 & 0.29 & 0.05 & 0.61 & 0.39 \\
                                                                  &&  Ghostery & 0.54 & 0.13 & 0.27 & 0.06 & 0.60 & 0.40\\
                                                          && Adblock Plus & 0.47 & 0.20 & 0.25 & 0.09 & 0.56 & 0.44\\
                                                                  && Privacy Badger & 0.23 & 0.44 & 0.11 & 0.22 & 0.45 & 0.55\\
                                                                  && Disconnect & 0.17 & 0.50 & 0.08 & 0.25 & 0.42 & 0.58\\
\hline
Sequential 7-gram & OCSVM & NoScript & 0.71 & 0.06 & 0.14 & 0.09 & 0.80 & 0.20\\
                                                                  &&  Ghostery & 0.67 & 0.10 & 0.15 & 0.08 & 0.75 & 0.25\\
                                                          && Adblock Plus & 0.62 & 0.15 & 0.11 & 0.13 & 0.75 & 0.25\\
                                                                  && Privacy Badger & 0.27 & 0.5 & 0.07 & 0.16 & 0.43 & 0.57\\
                                                                  && Disconnect & 0.19 & 0.58 & 0.06 & 0.17 & 0.36 & 0.64\\
\hline
Syntactic & PU & NoScript & 0.50 & 0.07 & 0.36 & 0.07 & 0.57 & 0.43 \\
                                                                  &&  Ghostery & 0.47 & 0.10 & 0.35 & 0.08 & 0.55 & 0.45\\
                                                          && Adblock Plus & 0.43 & 0.14 & 0.30 & 0.13 & 0.56 & 0.44\\
                                                                  && Privacy Badger & 0.18 & 0.38 & 0.15 & 0.28 & 0.46 & 0.54\\
                                                                  && Disconnect & 0.13 & 0.44 & 0.12 & 0.31 & 0.44 & 0.56\\
\hline
Sequential 7-gram & PU & NoScript & 0.70 & 0.05 & 0.16 & 0.09 & 0.79 & 0.21\\
                                                                  &&  Ghostery & 0.65 & 0.10 & 0.16 & 0.09 & 0.74 & 0.26\\
                                                          && Adblock Plus & 0.61 & 0.14 & 0.12 & 0.13 & 0.74 & 0.26\\
                                                                  && Privacy Badger & 0.18 & 0.57 & 0.07 & 0.18 & 0.36 & 0.64\\
                                                                  && Disconnect & 0.26 & 0.49 & 0.07 & 0.18 & 0.44 & 0.56\\
\hline
\end{tabular}
\end{center}

\caption{Agreement and disagreement in classification of tracking and functional {JavaScript programs} between our classifiers and PP-Tools on the wild dataset; \textcolor{ikgreen}{\protect\rule[0pt]{2mm}{2mm}}~agreement, \textcolor{ikred}{\protect\rule[0pt]{2mm}{2mm}}~disagreement; $T_p$ and $F_p$ represent {JavaScript programs} classified as tracking and functional, respectively, by the PP-Tool $p$, and $T_c$ and $F_c$ represent {JavaScript programs} classified as tracking and functional, respectively, by the classifier $c$.}

\label{tab:wds:agreement}
\end{table*}

\subsubsection{Performance of Classifiers}

From Table~\ref{tab:wds:agreement}, we first note that the two classifiers are similar in terms of their agreement/disagreement with PP-Tools, yielding very high disagreement ratio in the syntactic model (ranging from 39\% to 58\%) and significant disagreement in the sequential 7-gram model (20\% to 64\%). Overall, the two classifiers are relatively more in agreement with NoScript, followed closely by Ghostery and Adblock Plus. This agreement, however, is mostly in terms of classifying tracking {JavaScript programs}. 

In terms of agreement on functional {JavaScript programs}, the two classifiers are more in tune with Disconnect and Privacy Badger. These two PP-Tools however are the least in agreement with our classifiers in terms of tracking {JavaScript programs} (ranging between 13\% and 27\%). One possible explanation for this, in-line with previous research results~\cite{Nikiforakis:2012:}, is that Disconnect mostly blocks social plugins and buttons such as Facebook Likes and Twitter Follow. This leaves a host of other trackers allowed. The results for Privacy Badger are similar because it mainly blocks {JavaScript programs} that track users across multiple websites through cookies, which is routinely done by social widgets.

\subsubsection{Effect of Feature Models}

Table~\ref{tab:wds:agreement} also shows that the sequential 7-gram model of the two classifiers is more in agreement with NoScript, Ghostery and Adblock Plus as compared to the syntactic model, by around 20\%. However, the difference is nominal for Disconnect and Privacy Badger, with our classifiers agreeing more with these two PP-Tools in the syntactic model. In the following, we further analyse the observed disagreement by using the sequential 7-gram model of the one-class SVM as it showed superior results during our validation experiments (Section \ref{subsec:evaluation:ecgt}).

\subsection{Analysing Disagreements}
\label{subsec:analyzingdisagreement}

We delve into the set of {JavaScript programs} on which \textit{all} PP-Tools and our one-class SVM with sequential 7-grams classifier disagree. 
We are interested in the two facets of disagreement: {JavaScript programs} that our classifier considers tracking but all the PP-Tools consider 
functional, i.e., the set $ T_c \cap_p F_{p}$, and {JavaScript programs} that our classifier deems tracking while all PP-Tools consider functional, i.e., $F_c \cap_p T_p$.   

The number of {JavaScript programs} for which our classifier and \textit{all} PP-Tools are in disagreement is 9,071, representing a surprisingly 
high 6\% of the total number of {JavaScript programs} in the wild dataset. These {JavaScript programs} are split as 4,610 for $ T_c \cap_p F_{p}$ and 
4,461 for $F_c \cap_p T_p$. 
 
Inspecting these sets of disagreement would shed light on the main reasons for disagreement. 
Unfortunately, manually inspecting thousands of {JavaScript programs} (using the process used for producing our labelled dataset) is a tedious and time consuming process. 
We instead took a pragmatic approach, where we randomly sampled 
100 {JavaScript programs} each from the two sets of disagreement. We then manually inspected each JavaScript from the two samples 
and classified them as tracking or functional following the rules and methodology described in Section~\ref{subsec:lds}. 
The results of this manual process are shown in Table~\ref{tab:disagreement:evaluation:labelled}. 

Our classifier is correct in its labelling of 75 out of the 100 {JavaScript programs} it considered tracking. 
All these {JavaScript programs} are marked as functional by all the PP-Tools, implying that the PP-Tools are correct in 
labelling only 25 of these {JavaScript programs}. Moreover, our classifier correctly deemed 81 out of the 100 {JavaScript programs} 
as functional, implying that the PP-Tools correctly labelled only 19 of the {JavaScript programs} in the random 
sample. Note that these numbers should not be taken as reflecting the overall classification performance of our 
classifier, which was validated in Section~\ref{subsec:evaluation:ecgt}. These samples merely represent the corner cases of complete disagreement 
with \textit{all} other PP-tools. In other words, these numbers do not directly give us the true and false positive rates of our classification methodology. 

\begin{table}[!htb]
\begin{center}
\tabcolsep=0.11cm
\begin{tabular}{c|c|c|c|c}
\hline
\multirow{2}{*}{Disagreement} & \multirow{2}{*}{Total} &\multirow{2}{*}{Sample} & \multicolumn{2}{c}{Manual Labelling} \\ 
\cline{4-5}
&&& Tracking & Functional\\
\hline
$T_c \cap_p F_p$  &  4,610 & 100 & \cellcolor{ikgreen}75 & \cellcolor{ikred}25 \\
$F_c \cap_p T_p$  &  4,461 & 100 & \cellcolor{ikred}19 &  \cellcolor{ikgreen}81\\
\hline
\end{tabular}
\end{center}

\caption{Comparison of random samples of disagreement between our classifier and \textit{All} PP-Tools. \textcolor{ikgreen}{\protect\rule[0pt]{2mm}{2mm}}~manual labelling agrees with classifier and disagrees with PP-Tools, \textcolor{ikred}{\protect\rule[0pt]{2mm}{2mm}}~manual labelling disagrees with classifier and agrees with PP-Tools.}

\label{tab:disagreement:evaluation:labelled}
\end{table}

We first look at the 75 {JavaScript programs} correctly labelled as tracking by our classifier, and incorrectly considered as functional by the PP-Tools. 
Table~\ref{tab:tccapfp} in Appendix~\ref{subsection:ex_analyzing_disagreement} shows 10 representative {JavaScript programs} from this sample. We identify two typical reasons the PP-Tools miss blocking these
 {JavaScript programs}: 
 \begin{enumerate}
\item PP-Tools do not perform regular expression matching on the body of {JavaScript programs} to identify known trackers. Examples from these ``misses'' are 
{JavaScript programs} \#2 and \#3 in Table~\ref{tab:tccapfp}, which are allowed by all PP-Tools even though the referred domain \texttt{doubleclick.net} is included in their blacklists; this is because these {JavaScript programs} refer to this domain in their body, and the PP-Tools perform a regular expression match only on the URL of the JavaScript within the \texttt{script} tag. 
\item As expected PP-Tools are unable to block trackers that are not in the blacklist. An example is JavaScript \#9 in the table which we manually check to be a social widget allowing users to `like' comments on the webpage while tracking the user activity which is then transmitted to the first party domain. All PP-Tools miss this JavaScript because it does not belong to a popular social media domain. Similarly, JavaScript \#4 in the table belongs to a Russian tracking and advertising service domain \texttt{i-vengo.com}, but it is not in the blacklists of PP-Tools.
\end{enumerate}

Our classifier correctly marked these {JavaScript programs} as tracking as these scripts were syntactically and structurally similar to the tracking {JavaScript programs} used for training our classifiers. We stress that our classifiers do not need to know about all tracking scripts a priori; in fact, our classifiers are able to find new tracking scripts leveraging the syntactic and semantic similarity between known tracking scripts and previously unknown tracking {JavaScript programs}. 

Next, we look at the 81 {JavaScript programs} correctly labelled as functional by our classifier, and incorrectly considered as tracking by the PP-Tools. Table~\ref{tab:fccaptp} in Appendix~\ref{subsection:ex_analyzing_disagreement} shows 10 {JavaScript programs} from this sample. The predominant reason for the PP-Tools mistakenly blocking these {JavaScript programs} is because they belong to a tracking domain, even though the JavaScript itself performs a useful functionality. A typical example is JavaScript \#10 in the table, which fetches content from the first party domain \texttt{buzzfeed.com} without sending or collecting user information.

Lastly, we believe that the main reason our classifier misclassified 25 functional {JavaScript programs} and 19 tracking {JavaScript programs} is due to their structural similarity with representatives of the opposite class. For instance, the JavaScript \texttt{jquery.cookie.js} in the website \texttt{pnc.com} modifies cookies for this non-tracking domain. The PP-Tools rightly allow this JavaScript because \texttt{pnc.com} is not a tracking domain. But, due to the structural similarity of this JavaScript with {JavaScript programs} that modify cookies for tracking domains, our classifier deemed it as tracking. Similarly, our classifier misclassified the JavaScript \texttt{count.js} that gathers comment statistics on the website \texttt{listverse.com} and sends this information to the domain \texttt{disqus.com}, which is listed as a tracker by the PP-Tools. Our classifier misread this due to its similarity with {JavaScript programs} that maintain comments on a webpage but do not send this information through to third party trackers. For brevity, we do not enlist samples of these two categories of {JavaScript programs} misclassified by our classifier.

%% file: discussion.tex
In the following, we discuss possible uses and limitations of our approach.

\subsection{Possible Uses}

We envision at least two different uses of our technique: 

\subsubsection{Browser Extension}

A natural application of our technique is a client-based browser extension to evade trackers. We are currently developing a Firefox browser extension which extracts the JavaScript programs while a webpage is being loaded (prior to rendering) and calculates the similarities of the observed JavaScript programs against the training model 
which is kept locally. As discussed later in this section, we aim to periodically update the training model using a semi-supervised learning technique \cite{Basu:2004}. {We believe that such an extension is practical, as our current system classifies in the order of milliseconds per website.}

\subsubsection{Updating Blacklists and Whitelists}

Another possible use of our technique is to improve the accuracy of existing PP-Tools by updating their filtering lists. These tools could submit sets of JavaScript programs embedded in webpages (randomly chosen via a web scanner) to the classifier which would identify them as functional or tracking. The domains corresponding to the URLs linking these JavaScript programs can then be deemed tracking or tracking-free. The newly identified URLs can be used to update the PP-tool's blacklists (generally locally stored on the user browser) or to refine whitelists used by some tools such as NoScript. However, it is important to note that the generated URLs might lead to errors,
potential false positives and negatives, 
as the same domain may produce both functional and tracking JavaScript programs, and thus the decision to mark the domain tracking or tracking-free lacks the contextual information used by the classifier. One trivial example is \texttt{http://static.bbci.co.uk} domain which can be observed in both functional and tracking JavaScript programs. Another important caveat is that the blacklist based approach inherently does not filter in-page JavaScript programs, and as such the PP-Tools considered in this paper do not block their execution, with the exception of NoScript (which, as we have seen, aggressively blocks execution of JavaScript programs).

\subsection{Limitations and Possible Improvements}

\subsubsection{Classification Arms Race - Feature-exploit Case}

In principle, machine learning based detection is prone to exploits that introduce features absent from the training set~\cite{Curtsinger:2011}. In our case, the tracker could introduce some unique or rare piece of tracking code in the JavaScript program. Due to the uniqueness of the resulting JavaScript program, its feature set is unlikely to be present in the training model, and is therefore likely to go undetected. This can particularly be the case with non-pervasive ``hand-crafted'' trackers. However, as these pieces of code become ubiquitous and with periodic re-training of the classifier, this exploit can be circumvented. 

\subsubsection{Classification Arms Race - Code Obfuscation Case}
\label{subsub:code-obfus}

A tracker might also evade detection by obfuscating a tracking JavaScript program either by renaming it or by making changes to its code. In the first case, since our approach is based on code similarity of JavaScript programs, renaming does not affect the efficiency of our classifiers  (unlike blacklist based PP-Tools). In the latter case, the attacker might \emph{(i)} rename function or variable names, \emph{(ii)} add or remove whitespaces, \emph{(iii)} add zero-impact or dummy code, \emph{(iv)} or encode the JavaScript code~\cite{Xu:2012:POT}. 

In the semantic feature models, our classifiers are resilient against the first two types of JavaScript code obfuscation strategies. Appendix~\ref{appa:obfuscation} illustrates this via an example. On the other hand, the attacker might evade detection by applying the last two types of JavaScript code obfuscation techniques. We believe that this again presents the classical arms race issue, in which the trackers pay higher cost in trying to obfuscate their code to evade detection. Moreover, the obfuscation will need the additional guarantee of being detection proof as our machine learning techniques can re-learn newly introduced tracking JavaScript code (if enough trackers decide to obfuscate and the obfuscated JavaScript code still have structural similarities). It is also important to mention that blacklist based approaches still persist in spite of the fact that trackers can change their URLs to evade popular patterns. Note that despite the availability of obfuscation tools, our classifiers achieve higher efficiency and detect trackers missed by contemporary PP-Tools. We nevertheless regard this as a limitation and believe it to be an interesting area of future research.   


\subsubsection{The manual labelling challenge}
\label{subsec:mlabelling}

In this work, we used a relatively small set of labelled tracking and functional JavaScript programs ($2, 612$ to be precise). Our choice was dictated largely by the time consuming nature of the labelling process. 
Obviously, the performance of the classifier can be improved by increasing this number. One approach is to rely on crowdsourcing and recruit ``tech-savvy'' users as reviewers of a JavaScript program. This is not a trivial task as it requires considerable effort in providing guidelines (i.e., Table \ref{tab:GTable}) to the reviewers, a platform for interaction, and the need to resolve conflicts in labelling due to variable technical expertise of reviewers.

Alternatively, hybrid schemes 
such as semi-supervised learning~\cite{Basu:2004} can be used. The basic idea behind semi-supervised learning is to use unlabelled data to improve the training model that has been previously built using only labelled data. Semi-supervised learning has previously been successfully applied to real-time traffic classification problems~\cite{Erman:2007}, showing that automated identification of re-training points is possible. {Note that our model does not have to be re-trained as frequently as updating the blacklists of PP-Tools. Blacklists need to be frequently updated so that PP-Tools can keep track of new tracker URLs. In our case, re-training is only required in case new tracking JavaScript programs with unconventional code structure emerge. This is expected to occur far less frequently.}

%% file: relatedwork.tex
In recent years, there has been much research on privacy implications of web tracking \cite{Gill:2013, Balaprivacyleakage:2011, Krishnamurthy:2007, TranDLJ12, Acar:2014:WNF:, Jang:2010:, Krishnamurthy:2010, Nikiforakis:2013:CME}. For a recent and comprehensive survey on web tracking techniques and tracking prevention policies, see~\cite{mayer2012third}. 

Specific to automated detection of trackers, recently Gugelmann et al.~\cite{ref11} analysed HTTP traffic traces and identified statistical features of privacy-intrusive advertisements and analytics services. These features were extracted from HTTP requests sent from users' browsers to train machine-learning classifiers to detect privacy-intrusive services on the web. Their proposed scheme augments the blacklist of Adblock Plus by identifying previously unknown trackers. Our classification approach relies on the code structure of {JavaScript programs} instead of analysing HTTP request statistics which can be hard to gather and can have privacy issues of their own~\cite{ref11}. While our approach can be used as only a client-side (browser) service, their methodology requires collaboration and gathering of network traces that includes other services and nodes; a task difficult to do in a privacy-preserving manner. Lastly, the tool from~\cite{ref11} can only be used jointly with a blacklist based PP-Tool such as Adblock Plus. Our approach completely dismisses blacklists, and hence the need to manually update and maintain them frequently. 

Similarly, Orr et al.~\cite{Orr:2012} proposed a machine learning approach to detect JavaScript-based advertisements. They identify 20 key features of advertisement related {JavaScript programs} using static code analysis, and use SVM to achieve a classification accuracy of 98\%. Our work uses a much broader class of trackers in addition to advertisements including analytics, cookie readers/writers and social media widgets. Static code analysis to identify features from such a broad category of trackers is not feasible, and hence we use an approach that uses a dynamic feature space automatically extracted from code structure of {JavaScript programs}. Furthermore, Orr et al. do not compare the performance of their classifier against PP-Tools that specifically block advertisement. 

Tran et al. \cite{Tran:2012:TTF:} use JavaScript tainting to privacy leakage and to detect \emph{one} tracking website in addition to the services listed by Ghostery. In contrast, our framework reveals more than 4K new sites (i.e., JavaScript programs) showing tracker-like behavior, and based on a representative sample we estimate that 75\% indeed offer traditional advertisement and analytics services.

Our use of $n$-grams from program dependency graphs of {JavaScript programs} is adopted from the work of Hsiao, Cafarella and Narayanasamy \cite{Hsiao:2014:oopsala}, who demonstrated this technique to detect plagiarism in JavaScript code. However, they directly use $\mathsf{tf}\text{-}\mathsf{idf}$ scores of {JavaScript programs} to measure their similarities instead of using a machine learning approach. By comparison, the focus of our work is binary classification of functional and tracking {JavaScript programs}, and we use one-class machine learning algorithms on top of features extracted from $\mathsf{tf}\text{-}\mathsf{idf}$ scores of $n$-grams.

%% file: conclusion.tex
This paper presented a new automated mechanism to filter tracking {JavaScript programs} from functional {JavaScript programs} both embedded and externally linked within the DOM trees of webpages. 
We first study the (in)effectiveness of popular privacy-preserving tools with respect to balancing  
blocking of tracking {JavaScript programs} and allowing functional {JavaScript programs}. We then postulate and verify that 
one-class machine learning techniques that utilize similarities between tracking {JavaScript programs} based on syntactic and semantic features can effectively improve user's web experience.  One key aspect of our work is the ability of our classifiers to discover previously unseen tracking {JavaScript programs}. 
We stress that our methodology is generic and can be adapted to more conservative choices of what are considered functional JavaScript programs (e.g., JavaScript programs related to social media). Our current labelling is strict on what is considered tracking. Our classifiers can easily be tuned by simply adding or removing few instances of JavaScript programs from the tracking set used in training.

%% file: appendixa.tex
\subsection{PP-Tools Settings}
\label{appendix:pptoolsettings}
We used the following PP-Tools settings in our experiments to collect the dataset discussed in Section~\ref{subsec:obtain-js}.

\begin{table}[!ht]
\begin{center}

\begin{tabularx}{\linewidth}{ l l X  }
\toprule

PP-Tool & Filtering Method & Setting \\
\toprule
NS (v2.6.9.11) & Block all JS & Default blocking mode with iframes blocking option on \\
GT (v5.4.1) & Blacklist & Enabled `Blocking all tracker and cookies' option \\
AP (v2.6.7) & Blacklist & \emph{EasyList} and \emph{Fanboy's} list and disabled `Allow non-intrusive ads' option \\
DC (v3.14.0) & Blacklist & Default \\
PB (v0.1.4)  & Heuristics and & Default \\
   &  cookies blacklist  &  \\
\bottomrule
\end{tabularx}
\end{center}

\caption{PP-Tools' settings used with Firefox v32.0}
  \label{tab:pptoolssettings}
\end{table}

\subsection{Canonicalization of JavaScript programs}
\label{appendix:label:canonicalization}

The canonical form is an intermediate representation of the JavaScript program where variable names and loop specifics are abstracted away in an effort to apply natural language processing-like techniques on the resulting code
representation. To explain the canonical form and PDG of a JavaScript program, consider the following (toy) `equalTest' Javascript function:
\renewcommand{\lstlistingname}{Listing}
\setcounter{lstlisting}{0}
\lstdefinelanguage{MyScript}{
}
\lstset{
	tabsize=2,
	basicstyle=\footnotesize\ttfamily,
	showstringspaces=false,
	showspaces=false,
	language=MyScript
}
\begin{lstlisting}[caption={An Example of JavaScript Program},frame=single, label={list:toyexample}]
function equalTest(a, b){
	if(a == b){
		return true;}
	return false;}
\end{lstlisting}

\begin{figure}[!th]
\centering
\subfloat[\label{fig:pdg-equal-test}]{%
      \includegraphics[width=0.22\textwidth]{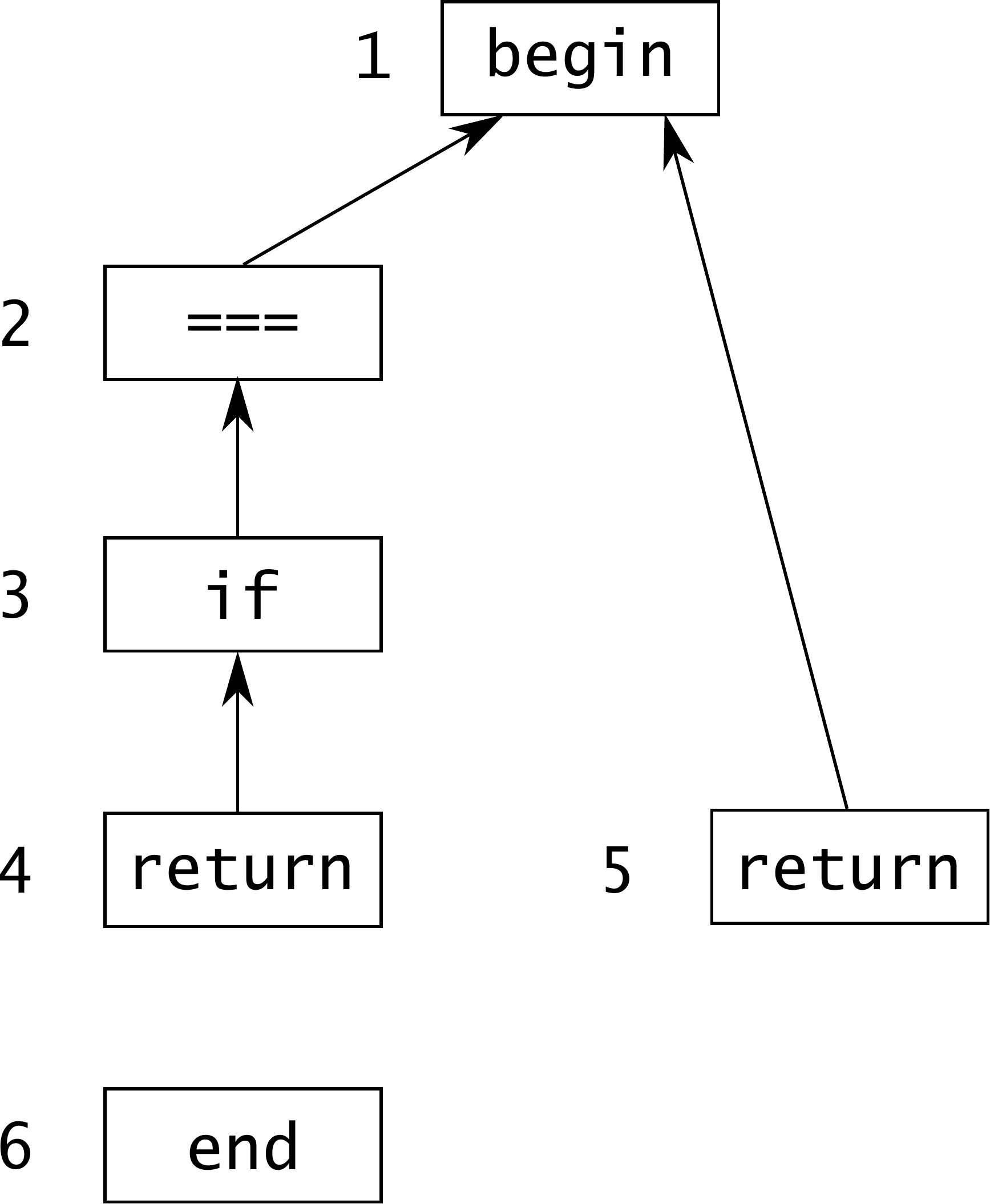}
}
\hfill
\subfloat[\label{fig:pdg-grams}]{%
      \includegraphics[width=0.16\textwidth]{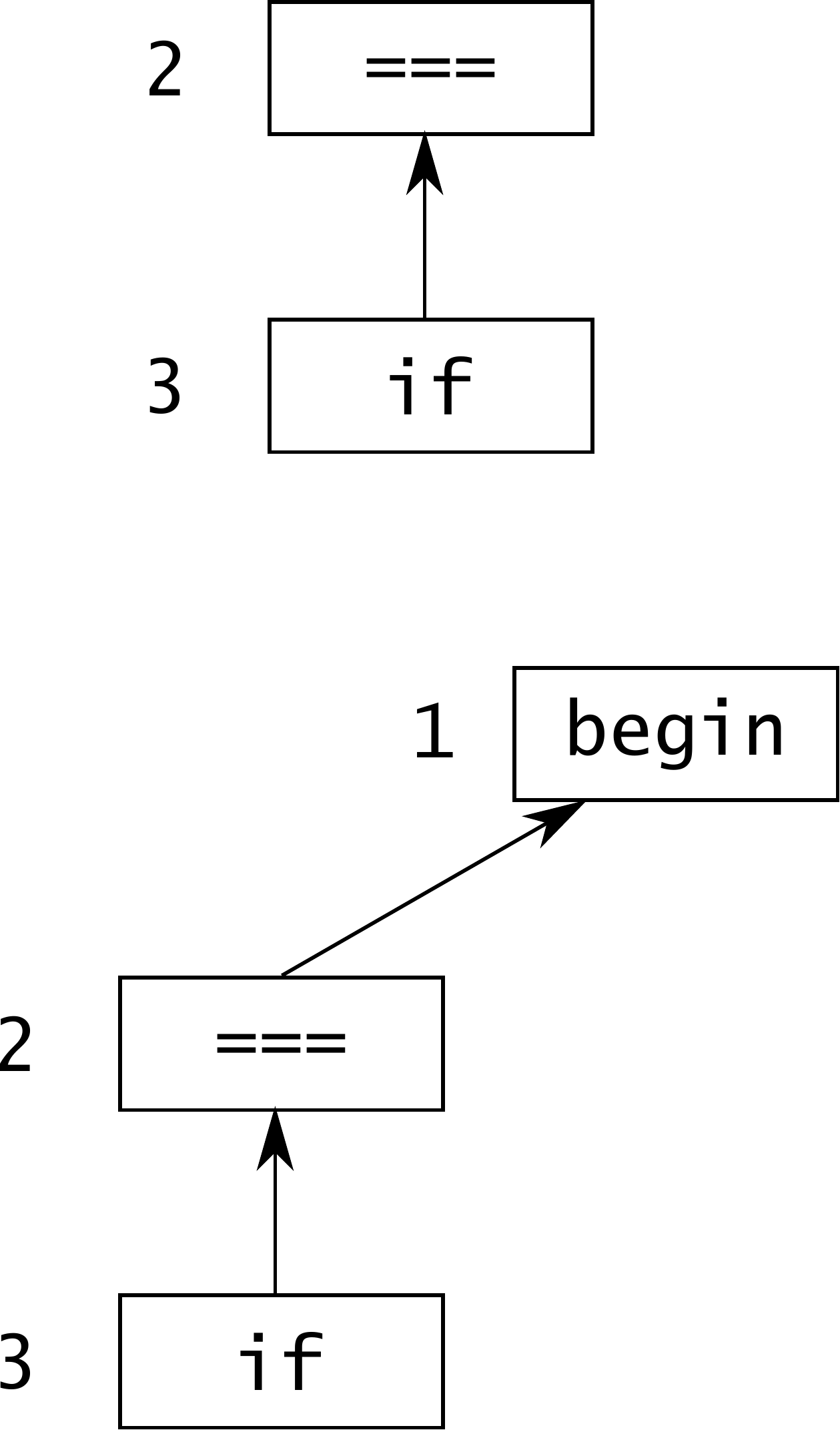}
}
\caption{
\protect\subref{fig:pdg-equal-test} 
Program dependency graph of the \texttt{equalTest} canonical form. 
\protect\subref{fig:pdg-grams}
 $2$-gram (top) and $3$-gram (bottom) of line 3 of the program dependency graph.} 
\end{figure}

The canonical form of this routine is:

\begin{lstlisting} [caption={Canonical form of JavaScript code in Listing \# 3}, frame=single, label={list:canonicalform}] 
function equalTest(a, b){
1:		begin;
2:		$0 = a === b;
3: 		if($0){
4:			return true;}
5:		return false;
6:		end;}
\end{lstlisting}

One line of the canonical form consists of a binary operation, its operands or an assignment. The PDG of this routine is shown in Figure~\ref{fig:pdg-equal-test}. The $2$ and $3$-grams of line 3 above is shown in Figure~\ref{fig:pdg-grams}. For a more detailed example of these concepts, see~\cite{Hsiao:2014:oopsala}.

\subsection{Analyzing Disagreement : Examples}
\label{subsection:ex_analyzing_disagreement}
\begin{figure*}[!ht]
\tabcolsep=0.11cm
\centering
\subfloat[Syntactic OCSVM.]{\label{fig:ocsvm:syn:agreement} 
      \includegraphics[width=0.240\textwidth, height=0.30\textwidth, keepaspectratio]{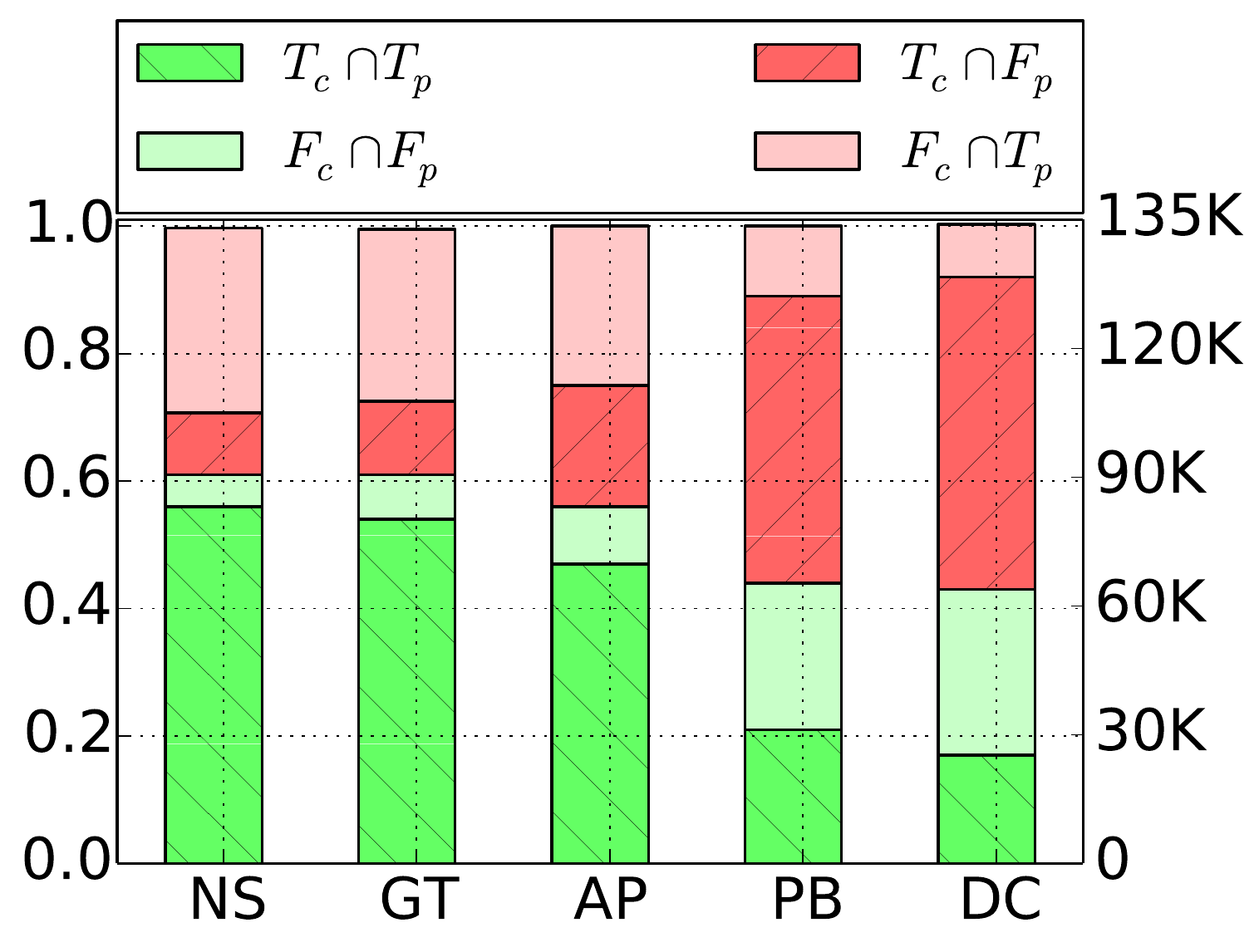}
      }
      \subfloat[Sequential 7-gram OCSVM.] {\label{fig:ocsvm:seq7:agreement} 
      \includegraphics[width=0.240\textwidth, height=0.30\textwidth, keepaspectratio]{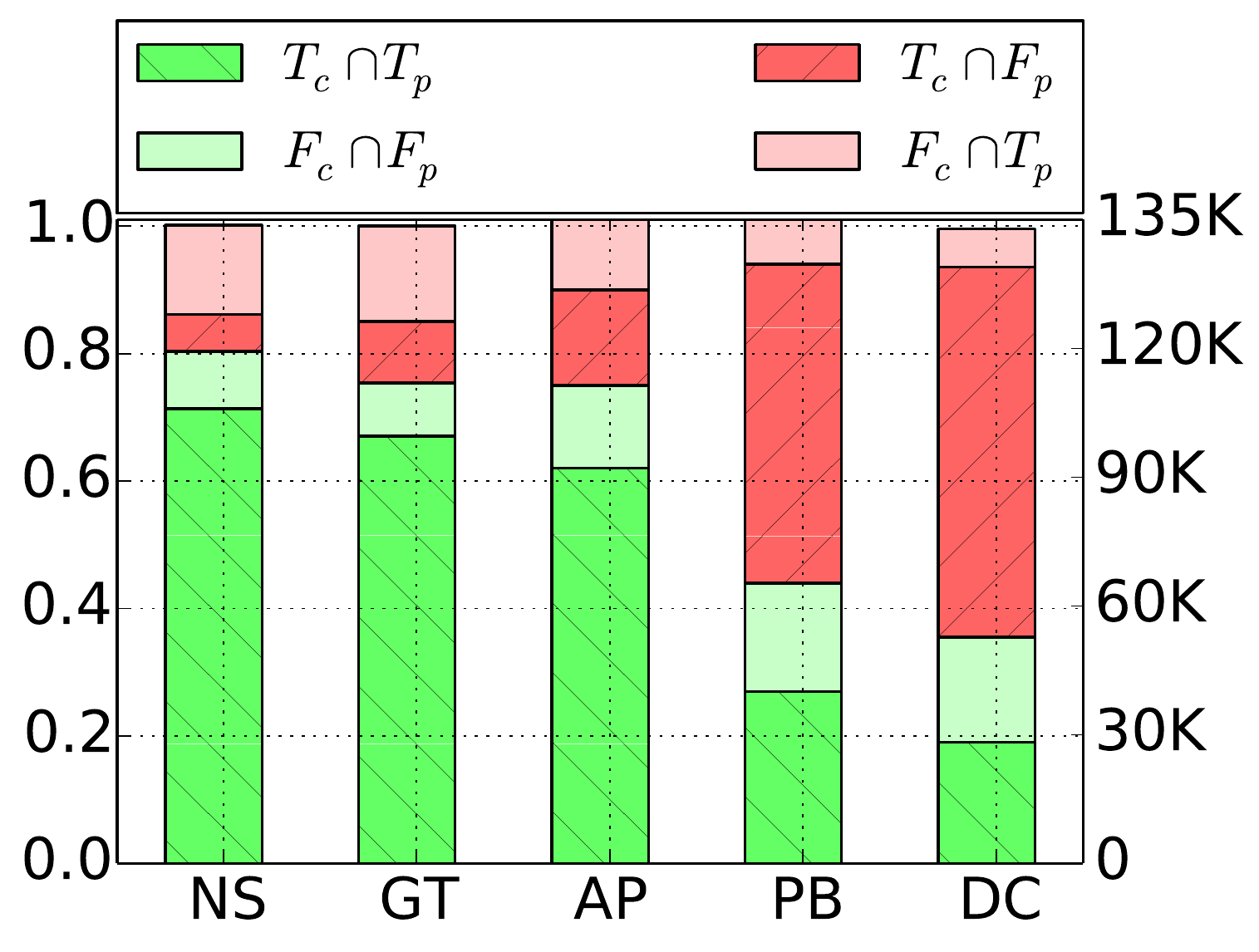}
      }
\subfloat[Syntactic PU.]{\label{fig:pu:syn:agreement} 

\includegraphics[width=0.240\textwidth, height=0.30\textwidth, keepaspectratio]{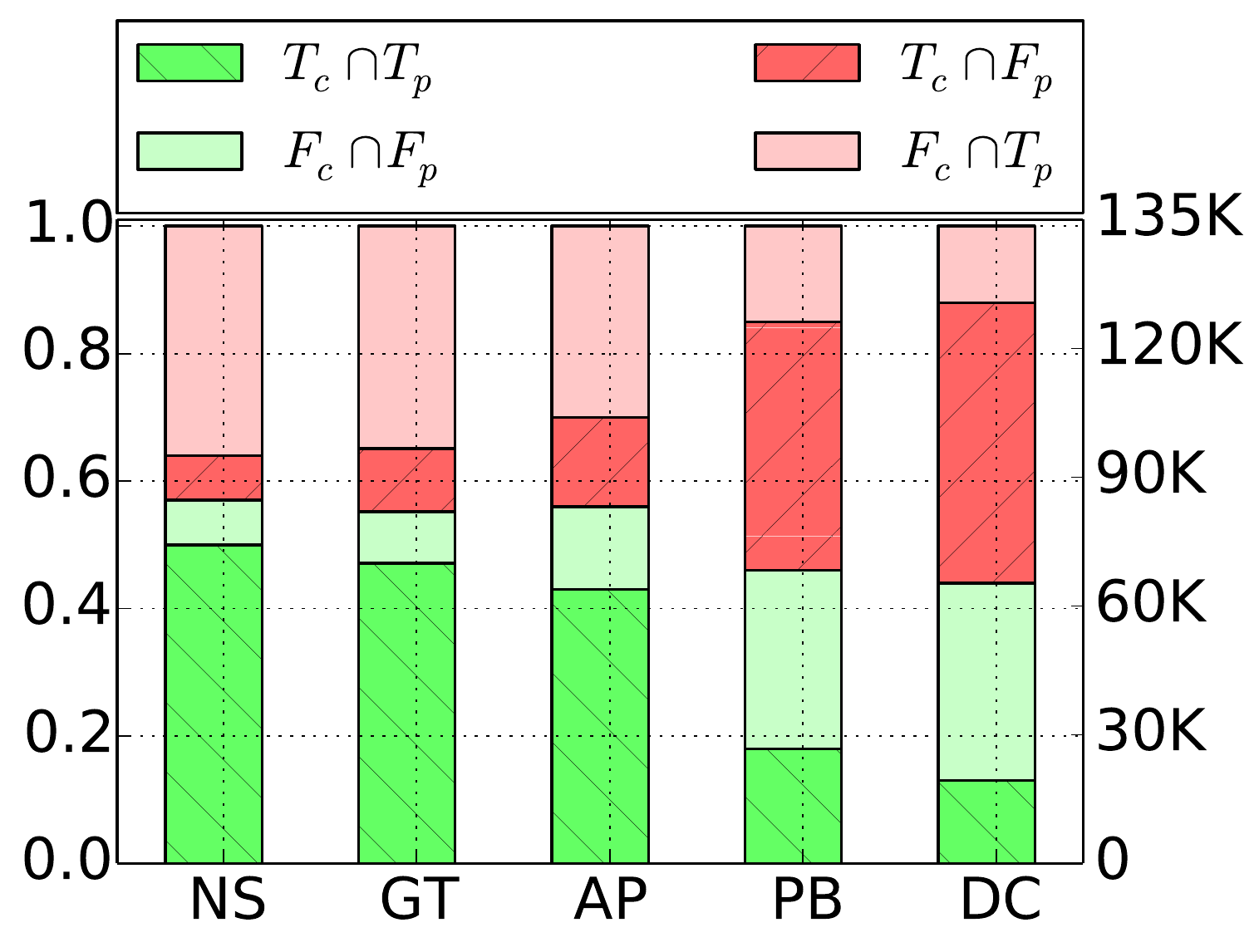}
}
      \subfloat[Sequential 7-gram PU.] {\label{fig:pu:seq7:agreement}

            \includegraphics[width=0.240\textwidth, height=0.30\textwidth, keepaspectratio]{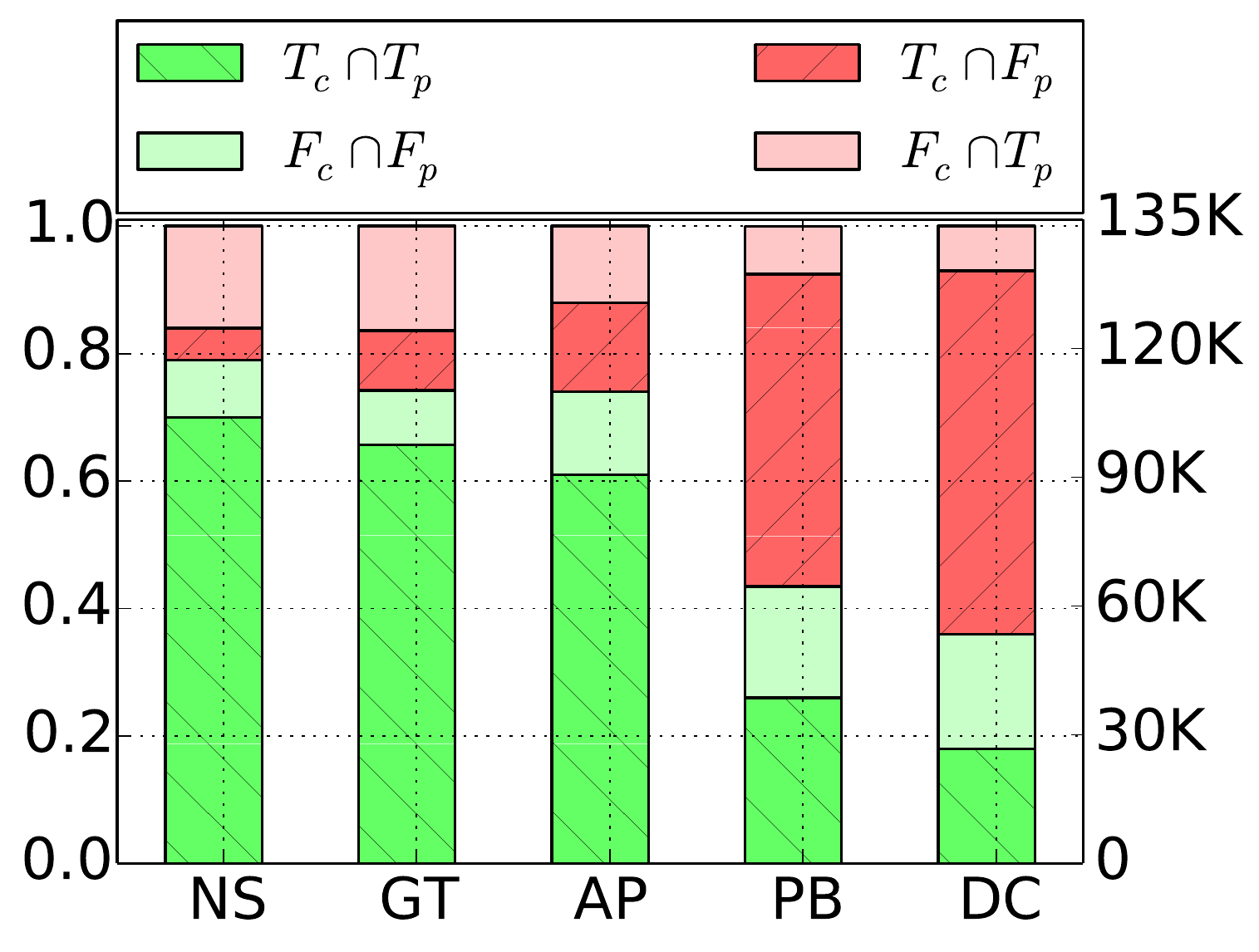}
}
\caption{Agreement (\textcolor{ikdarkgreen}{\protect\rule[0pt]{2mm}{2mm}} \& \textcolor{ikgreen}{\protect\rule[0pt]{2mm}{2mm}}) and disagreement (\textcolor{ikdarkred}{\protect\rule[0pt]{2mm}{2mm}} \& \textcolor{ikred}{\protect\rule[0pt]{2mm}{2mm}}) in classification of tracking and functional {JavaScript programs} between our classifiers and PP-Tools on the wild dataset. $T_p$ and $F_p$ represent {JavaScript programs} classified as tracking and functional, respectively, by the PP-Tool $p$, and $T_c$ and $F_c$ represent {JavaScript programs} classified as tracking and functional, respectively, by the classifier $c$; NS, GT, AP, PB, and DC stand for NoScript, Ghostery, Adblock Plus, Privacy Badger, and Disconnect, respectively.}
\label{fig:ocsvm:agreement}
\end{figure*}

Examples where there is disagreement between our classifier and PP-Tools are shown in Table~\ref{tab:tccapfp} and Table \ref{tab:fccaptp}. In essence, we show examples of JavaScript codes that are classified as tracking JavaScript codes (resp. functional) while classified as functional (resp. tracking) by the ensemble of PP-Tools. We discuss these details in Section~\ref{subsec:analyzingdisagreement}. For reference, the ratio of agreement and disagreement between our classifier and PP-Tools on the wild dataset is also illustrated in Figure~\ref{fig:ocsvm:agreement}. 

\begin{table*}[h!]
\tabcolsep=0.11cm
\begin{center}
\begin{tabularx}{\linewidth}{ c l X l l}
\hline
\# & Website & JavaScript Program & Referred Domain & Function Performed\\ \midrule 

1 & examiner.com & cdn2-b.examiner.com/.../ex$\_$omniture/s$\_$code.js & omniture.com & Analytics\\
2 & bbc.com & static.bbci.co.uk/bbcdotcom/.../adverts.js & pubads.g.doubleclick.net & Analytics + Ads\\
3 & telegraph.co.uk & telegraph.co.uk/template/ver1-0/js/gpt.js & pubads.g.doubleclick.net & Analytics + Ads\\
4 & vesti.ru & s.i-vengo.com/js/ivengo.min.js & www.i-vengo.com & Analytics + Ads\\
5 & climatempo.com.br & http://s1.trrsf.com/metrics/inc/br/201411250000d.js & scorecardresearch.com & Analytics \\
6 & amc.com & amc.com/wp-content/plugins/amcn-common-analytics/js/common-analytics.js & omniture.com & Track user activities \\
7 & lancer.com & static.lancers.jp/js/ga\_social\_tracking.js & google.com & Tracker user activities \\
8 & iqiyi.com & static.iqiyi.com/js/pingback/qa.js & pps.tv, baidu.com, 71.com & Tracker user activities \\
9 & babyblog.ru & act.babyblog.ru/static844/likes.js & babyblog.ru & Social widgets \\
10 & autoscout.de & s.autoscout24.net/unifiedtracking/gtm.js & autoscout.de & Tracks user activities \\

\hline
\end{tabularx}
\end{center}

\caption{Ten {JavaScript programs} our classifier correctly classified as tracking and all the PP-Tools wrongly classified as functional verified through manual labelling.}

\label{tab:tccapfp}
\end{table*}

\begin{table*}[h!]
\begin{center}
\tabcolsep=0.11cm
\begin{tabularx}{\linewidth}{ c X l l X}
\hline
\# & Website & JavaScript Program & Referred Domain & Function Performed\\ \midrule 

1 & crateandbarrel.com & j.c-b.co/js/account$\_$1505080410.js & crateandbarrel.com & Creates user accounts  \\
2 & suomi24.fi & kiwi27.leiki.com/focus/mwidget.js & leiki.com & Magnifying widget \\
3 & nhl.com	 & b3.mookie1.com/2/LB/3115965742.js	& cdn-akamai.mookie1.com	& Fetches content \\
4 & michael.com & edgesuite.net/js/picturefill.min.js & - & Fetches content \\
5 & ing.nl & ensighten.com/ing/NL-ingnl-prod/code/fc$\_$90aaa8fc7.js & ing.nl & Fetches content \\
6 & worldoftanks.ru & mc.wargaming.net/tsweb.js & wargaming.net & Sets session cookie \\
7 & divyabhaskar.co.in & nr.taboola.com/newsroom/bhaskar-divyabhaskar/getaction.js & taboola.com & Enables user interaction\\
8 & 15min.lt & 15minadlt.hit.gemius.pl/$\_$1431091788674/redot.js & squarespace.com & Timestamps user login \\
9 & abovetopsecret.com & casalemedia.com/j.js & abovetorespect.com & Fetches content \\
10 & buzzfeed.com & ct-ak.buzzfeed.com/wd/UserWidget.js & s3.amazonaws.com & Fetches content\\ 
\hline
\end{tabularx}
\end{center}
\caption{Ten {JavaScript programs} our classifier correctly classified as functional and all the PP-Tools incorrectly classified as tracking verified through manual labelling.}
\label{tab:fccaptp}
\end{table*}

\vspace{+0.15in}
\subsection{JavaScript Code Obfuscation}
\label{appa:obfuscation}
Criteo\footnote{\url{http://www.criteo.com}} sets its tracking cookies at \texttt{dailymotion.com} with the JavaScript code shown in Listing~\ref{list:obfexample}. By using an online obfuscation tool,\footnote{\url{http://www.danstools.com/javascript-obfuscate/}} the obfuscated version of this code is shown in Listing~\ref{list:obfuscationcombined}, with variable names replaced and white spaces removed, i.e., type \emph{(i)} and \emph{(ii)} obfuscation mentioned in Section~\ref{subsub:code-obfus}. The canonical form of both the original code and the obfuscated code is the same, as shown in Listing~\ref{list:Obfuscatedcanonical}, meaning that our semantic feature models based classifiers will still detect the obfuscated trackers. 

\begin{lstlisting}[caption={An Example of JavaScript Program},frame=single, label={list:obfexample}]
var crtg_nid="1822";
var crtg_cookiename="co_au";
var crtg_varname="crtg_content";
function crtg_getCookie(c_name){
 var i,x,y,ARRCookies=document.cookie.split(";");
 for(i=0;i<ARRCookies.length;i++){
 x=ARRCookies[i].substr(0,ARRCookies[i]
          .indexOf("="));
 y=ARRCookies[i].substr(ARRCookies[i]
          .indexOf("=")+1);
 x=x.replace(/^\s+|\s+$/g,"");
 if(x==c_name){
	return unescape(y);
	}}
 return "";
}
\end{lstlisting}
\lstset{
	tabsize=2,
	basicstyle=\footnotesize\ttfamily,
	showstringspaces=false,
	showspaces=false,
	language=MyScript
}

\begin{lstlisting}[caption={Type \emph{(i)} and \emph{(ii)} obfuscation.}, frame=single, label={list:obfuscationcombined}]
var a="1822";var b="co_au";var c="crtg_
content";function crtg_getCookie(e){var i,x,y,
d=document.cookie.split(";");while(i<d.length)
{x=d[i].substr(0,d[i].indexOf("="));y=d[i].sub
str(d[i].indexOf("=")+1);x=x.replace(/^\s+|\s+
$/g,"");if(x==e){return unescape(y)}i++}
return""}
\end{lstlisting}

\begin{lstlisting}[caption={Canoncial form of original and obfuscated code.}, frame=single, label={list:Obfuscatedcanonical}]
function crtg_getCookie = 
             function crtg_getCookie(e)
{begin;$0 = document.cookie;
	d = $0.split(";");
	$1 = d.length; $2 = i < $1;
	while ($2) {
		$3 = d[i];$4 = d[i];
		$5 = $4.indexOf("=");
		x = $3.substr(0, $5);
		$6 = d[i];$7 = d[i];
		$8 = $7.indexOf("=");
		$9 = $8 + 1;
		y = $6.substr($9);
		x = x.replace(RegExp("^\s+|\s+$","g"), "");
		$10 = x == e;
		if ($10) {
			$11 = unescape(y);
			return $11;}
		$12 = i; i = i + 1;
		$13 = d.length;
		$2 = i < $13;}
	return "";
	end;};
  $14 = %InitializeVarGlobal("a", 0, "1822");
  $15 = %InitializeVarGlobal("b", 0, "co_au");
  $16 = %InitializeVarGlobal("c", 0, 
               "crtg_content");
  end;
\end{lstlisting}

\subsection{Surrogate JavaScript Programs}
\label{sec:appendixb}

While using PP-Tools, certain content might not be working properly~\cite{brokenpages}. This is known as \textit{broken web-pages}. This happens when certain web-components are blocked on a website which might be necessary for smooth browsing. In order to tackle broken pages, exceptions and errors, PP-Tools often inject snippets of non-tracking JavaScript programs, also called \textit{surrogate} scripts, when they block content from loading. Through manual inspection, we observed that both Ghostery and NoScript inject surrogate scripts. We investigated Ghostery and NoScript source codes to derive a comprehensive list of {surrogates}. 

\begin{figure}[!th]
\centering
\subfloat[\label{fig:surrogategh}]{%
      \includegraphics[width=0.24\textwidth, height=0.30\textwidth, keepaspectratio]{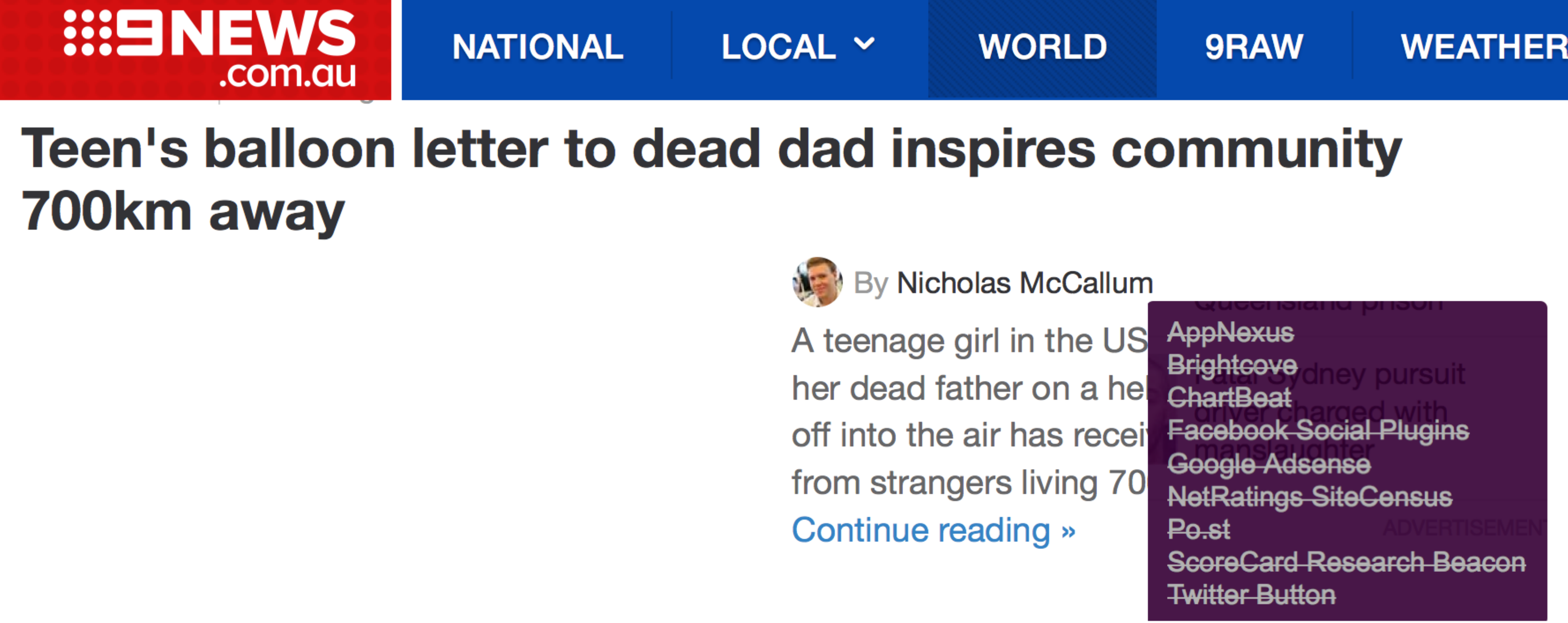}
      }
      \subfloat[\label{fig:surrogatewtgh}]{%
      \includegraphics[width=0.24\textwidth, height=0.30\textwidth, keepaspectratio]{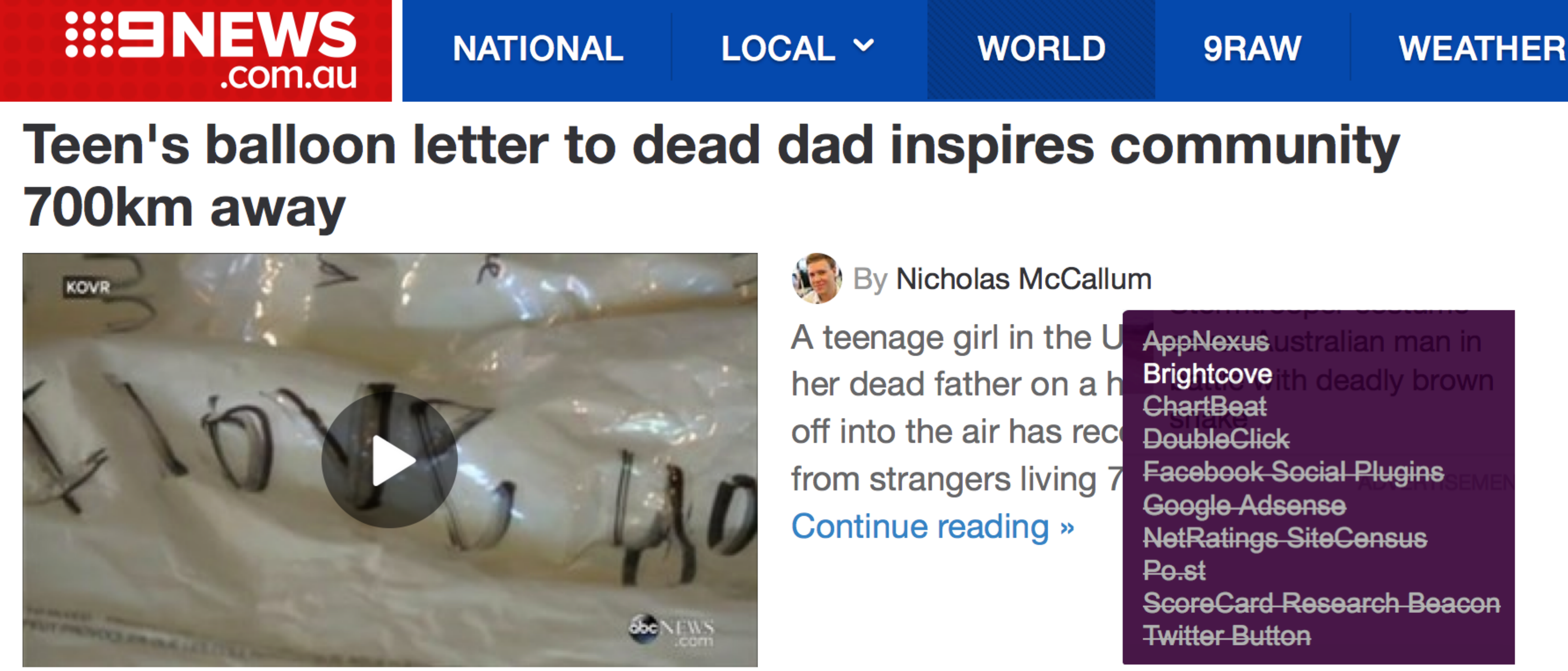}
      }

\caption{Ghostery's filters \protect\subref{fig:surrogategh} video content by blocking \texttt{brightcove} widget on www.9news.com.au. Once unblocked, \texttt{brightcove} loads \protect\subref{fig:surrogatewtgh} video content on the web-page.}
\label{fig:surrogate-example}
\end{figure}

Interestingly, however, we noticed that using its `block all trackers' setting, certain Ghostery surrogate scripts do not necessarily facilitate smooth browsing. Instead, they block useful content. For instance, Figure~\ref{fig:surrogate-example} shows an example where Ghostery injects a surrogate script 
 for the \texttt{brightcove} widget. The resulting surrogate script blocks the video content on the webpage, which is arguably a useful functionality. 
\subsection{JavaScript Program Similarity}
\label{appendix:jssim}
In this section, we present examples of JavaScript programs in the wild dataset that are labelled as tracking by our classifier. Our aim is to show how our classifier trained on the labelled dataset was successful in finding tracking JavaScript codes that were absent from the training set. The JavaScript programs in listing~\ref{list:simExample3} and listing~\ref{list:simExample4} from the test set were marked as tracking by our classifier. These two were not present in the training set. Two JavaScript programs that were present in the training data are shown in listing~\ref{list:simExample1} and listing \ref{list:simExample2}. We find that these two have normalized cosine similarity values of 0.89 and 0.73 with JavaScript programs in listing~\ref{list:simExample3} and listing~\ref{list:simExample4}, respectively. On the other hand the JavaScript programs in listing~\ref{list:simExample3} and listing~\ref{list:simExample4} had cosine similarities of $\leq$ 0.17 and $\leq$ 0.21, respectively, with all functional JavaScripts (cf. \S~\ref{subsec:fselection}). This suggests that these JavaScript programs are distinct from functional JavaScript programs while being similar to tracking JavaScript programs, the key idea exploited in our approach. 

\begin{lstlisting} [caption={ClickTale-Behavioural Tracker JavaScript Code snippet on cnn.com. It writes and reads cookie to track user on cnn.com.}, frame=single, label={list:simExample1}] 
createCookie: function (name,value,days) 
{
 if (days)
 {
  var date = new Date();
  date.setTime(date.getTime( )+( days*24*60*
  60*1000));
  var expires = "; expires="+date.toGMTString( );
 }
else var expires = "";
  document.cookie = name+"="+value+expires+"; 
  path=/";
},
readCookie : function (name) 
{
  var nameEQ = name + "=";
  var ca = document.cookie.split( ';');
  for( var i=0;i < ca.length;i++) {
  var c = ca[i];
  while ( c.charAt( 0)==' ') c = c.substring( 1,
  c.length);
  if ( c.indexOf( nameEQ) == 0) 
  return c.substring( nameEQ.length,c.length);
 }
  return null;
  }
}
\end{lstlisting}
\vspace{-0.10in}
\begin{lstlisting} [caption={Cookie setting JavaScript code on bbc.com. The code snippet sets Google analytic cookies to track on bbc.com.}, frame=single, label={list:simExample2}] 
  var _gaq = _gaq || [];
  _gaq.push(['_setAccount', 'UA-1627489-1']);
  _gaq.push(['_setDomainName', 'bbc.com']);
  _gaq.push(['_trackPageview']);
  (function () 
  {
  var ga = document.createElement('script'); 
  ga.type = 'text/javascript'; 
  ga.async   = true;
  ga.src = ('https:' == document.location.protocol
   ? 'https://ssl' : 'http://www') + 
   '.google-analytics.com/ga.js';
  var s = document.getElementsByTagName
  ('script')[0];
  s.parentNode.insertBefore(ga, s);
})(); 
\end{lstlisting}
\vspace{-0.10in}
\begin{lstlisting} [caption={The code excerpt from browser-detection.js on geo.tv. The code writes and reads cookies and finger-prints user's browser on geo.tv.}, frame=single, label={list:simExample3}] 
writeCookie: function(name, value, days){
var expiration = ""; 
if(parseInt(days) > 0)
{
  var date = new Date();
  date.setTime(date.getTime() + parseInt(days) 
  * 24 * 60 * 60 * 1000);
  expiration = '; expires=' + date.toGMTString();
}
  document.cookie = name + '=' + value + 
  expiration + '; path=/';
},
readCookie: function(name){
 if(!document.cookie){ return ''; }		
 var searchName = name + '='; 
 var data = document.cookie.split(';');
 for(var i = 0; i < data.length; i++){
 while(data[i].charAt(0) == ' '){
   data[i] = data[i].substring(1, 
   data[i].length);
 }
 if(data[i].indexOf(searchName) == 0){ 
   return data[i].substring(searchName.length, 
   data[i].length);
 }
}
 return '';
}
\end{lstlisting}
\vspace{-0.10in}
\begin{lstlisting} [caption={Cookie setting JavaScript code on dailymotion.com. The code sets Marin Software's (marinsoftware.com) tracking and advertisement cookie on dailymotion.com.}, frame=single, label={list:simExample4}] 
var _mTrack = _mTrack || []; 
_mTrack.push(['trackPage']); 
(function() {
  var mClientId = 'f52zgtujz0';
  var mProto = ('https:' == document.location
  .protocol ? 'https://' : 'http://'); 
  var mHost = 'tracker.marinsm.com';
  var mt = document.createElement('script'); 
  mt.type = 'text/javascript';
  mt.async = true; mt.src = mProto + mHost + 
  '/tracker/async/' + mClientId + '.js';
  var fscr = document.getElementsByTagName
  ('script')[0];     
  fscr.parentNode.insertBefore(mt, fscr);
})();
\end{lstlisting}

In the future, we aim to characterize and study JavaScript code obfuscation techniques employed by trackers. In essence, we aim to investigate JavaScript's dynamic code generation and run-time evaluation functions, e.g., \texttt{eval()} and \texttt{document.write()}.
We believe that the investigation of the arguments supplied to these functions can be leveraged in the re-training of our classifiers and in the detection of possible obfuscated tracking and malicious JavaScript programs.